\documentclass[12pt]{article}
\usepackage{fancybox}
\usepackage[dvips]{graphicx}

\newcommand {\beq}{\begin{equation}}
\newcommand {\eeq}{\end{equation}}
\newcommand {\beqa}{\begin{eqnarray}}
\newcommand {\eeqa}{\end{eqnarray}}
\newcommand {\beqan}{\begin{eqnarray*}}
\newcommand {\eeqan}{\end{eqnarray*}}
\newcommand {\n}{\nonumber \\}

\makeatletter
\@addtoreset{equation}{section}

\makeatother

\begin{document}
\setlength{\oddsidemargin}{0cm}
\setlength{\baselineskip}{7mm}

\begin{titlepage}
 \renewcommand{\thefootnote}{\fnsymbol{footnote}}
$\mbox{ }$
%\vspace{-3cm}
\begin{flushright}
\begin{tabular}{l}
KEK-TH-748 \\
%{\bf KEK preprint 2001 }\\
KUNS-1708\\
hep-th/0102168\\
Feb 2001
\end{tabular}
\end{flushright}

~~\\
~~\\
~~\\

\vspace*{0cm}
    \begin{Large}
       \vspace{2cm}
       \begin{center}
         {Supermatrix Models}      \\
       \end{center}
    \end{Large}

  \vspace{1cm}

\begin{center}
           Takehiro A{\sc zuma}$^{1)}$\footnote
           {
e-mail address : azuma@gauge.scphys.kyoto-u.ac.jp},
           Satoshi I{\sc so}$^{2)}$\footnote
           {
e-mail address : satoshi.iso@kek.jp},
           Hikaru K{\sc awai}$^{1)}$\footnote
           {
e-mail address : hkawai@gauge.scphys.kyoto-u.ac.jp} \\
and
           Yuhi O{\sc hwashi}$^{1),}$$^{3)}$\footnote
           {
e-mail address :  yuhi@gauge.scphys.kyoto-u.ac.jp} \\
       $^{1)}$ {\it Department of Physics, Kyoto University,
Kyoto 606-8502, Japan}\\
        $^{2)}$ {\it High Energy Accelerator Research Organization (KEK),}\\
               {\it 1-1Oho, Tsukuba, Ibaraki 305-0801, Japan} \\
       $^{3)}$ {\it Department of Accelerator Science,}\\
        {\it The Graduate University for Advanced Studies,}\\
        {\it 1-1Oho, Tsukuba, Ibaraki 305-0801, Japan}\\
\end{center}

\vfill

\begin{abstract}
\noindent
We investigate several matrix models based on super Lie algebras,
$osp(1|32,R)$, $u(1|16,16)$ and $gl(1|32,R)$.
They are natural generalizations of IIB matrix model and
were first proposed by Smolin \cite{Smolin:2000kc}.
In particular, we study the supersymmetry structures of these models
and discuss  possible reductions to IIB matrix model.
We also point out that diffeomorphism invariance is hidden
in gauge theories on noncommutative space which are derived from matrix
 models. This symmetry is independent of the global $SO(9,1)$
 invariance in IIB matrix model
and we report our trial to extend the global Lorentz invariance to 
local symmetry
by introducing $u(1|16,16)$ or $gl(1|32,R)$ super Lie algebras.

\end{abstract}
\vfill
\end{titlepage}
\vfil\eject

\section{Introduction}
\setcounter{equation}{0}
A large $N$ reduced model has been proposed as a nonperturbative
formulation of type IIB superstring theory\cite{IKKT}\cite{FKKT}.
It is defined by the following action:
\beq
S  =  -{1\over g^2}Tr({1\over 4}[A_{\mu},A_{\nu}][A^{\mu},A^{\nu}]
+{1\over 2}\bar{\psi}\Gamma ^{\mu}[A_{\mu},\psi ]) .
\label{action}
\eeq
It is a large $N$ reduced model \cite{RM} of 10-dimensional ${\cal N}=1$
super Yang-Mills theory.
Here $\psi$ is a 10-dimensional Majorana-Weyl spinor field, and
$A_{\mu}$ and $\psi$ are $N \times N$ Hermitian matrices.
This model is called IIB or IKKT matrix model.
It is formulated in a manifestly covariant way which enables us
to study the nonperturbative issues of superstring theory.
In fact we can in principle predict the dimensionality of spacetime,
the gauge group and the matter contents by solving this model.
Such possibilities have been discussed
in \cite{AIKKT}\cite{Nishimura}.
We refer a review for more
detailed expositions and references\cite{review}.

Although we have not yet obtained the complete interpretation of the
model as the theory of gravity,
the following arguments on supersymmetry
lead us to  interpret distributed eigenvalues as the extent of space-time.
In addition to the original supersymmetry of the ${\cal N}=1$
super Yang-Mills
\beqa
\delta^{(1)}\psi &=& \frac{i}{2}
                     [A_{\mu},A_{\nu}]\Gamma^{\mu\nu}\epsilon ,\n
\delta^{(1)} A_{\mu} &=& i\bar{\epsilon }\Gamma^{\mu}\psi ,
\label{Ssym1}
\eeqa
the reduced model action (\ref{action}) is invariant under the
following shift of the fermionic variable
\beqa
\delta^{(2)}\psi &=& \xi ,\n
\delta^{(2)} A_{\mu} &=& 0.
\label{Ssym2}
\eeqa
Since the original 10-dimensional space-time is reduced to a single
point, only the  repetitions of the first transformations can no longer
generate space-time translation. (It vanishes up to $SU(N)$ transformation.)
However,
if we take a linear combination of $\delta^{(1)}$ and $\delta^{(2)}$ as
\beqa
\tilde{\delta}^{(1)}&=&\delta^{(1)}+\delta^{(2)}, \n
\tilde{\delta}^{(2)}&=&i(\delta^{(1)}-\delta^{(2)}),
\eeqa
we obtain an enhanced $\cal{N}$=2 supersymmetry algebra
\beqa
(\tilde{\delta}^{(i)}_{\epsilon}\tilde{\delta}^{(j)}_{\xi}
    -\tilde{\delta}^{(j)}_{\xi}\tilde{\delta}^{(i)}_{\epsilon})\psi   &=&0 ,\n
(\tilde{\delta}^{(i)}_{\epsilon}\tilde{\delta}^{(j)}_{\xi}
    -\tilde{\delta}^{(j)}_{\xi}\tilde{\delta}^{(i)}_{\epsilon})A_{\mu}&=&
                                - 2i\bar{\epsilon}\Gamma^{\mu}\xi
\delta_{ij}.
\eeqa
The r.h.s. is a shift of the bosonic variables
\beq
\delta A^{\mu}  = c^{\mu},
\eeq
where $c_{\mu}$ is proportional to unit matrix.
The reduced model action (\ref{action}) is, 
of course, invariant under this shift.
Hence, if we interpret the bosonic variables as space-time
coordinates,  the above ${\cal N}=2$ supersymmetry generates
translation in the new space-time.

\par
Another reason to consider the bosonic variables $A^{\mu}$ as the space-time
comes from  the relation between matrix models and field theory on
noncommutative geometry.
As have been investigated  in papers \cite{NCYM}, matrix models can
be rewritten as gauge theories on noncommutative space by expanding
the bosonic variables $A^{\mu}$ around the noncommutative background
$\hat{x}_{\mu}$ satisfying
\begin{equation}
 [\hat{x}^{\mu},  \hat{x}^{\nu}] = - i \theta^{\mu \nu},
\label{noncommutative-x}
\end{equation}
where  $\theta^{\mu\nu}$ are $c$-numbers.
We assume the rank of $\theta^{\mu\nu}$
to be $\tilde{d}$ and define its inverse $\beta_{\mu\nu}$ in
$\tilde{d}$-dimensional  subspace.
$\hat{x}^{\mu}$ satisfy the canonical commutation relations and
they span the $\tilde{d}$-dimensional phase space.
The semiclassical correspondence shows that the
volume of the phase space (measured in the coordinate space of
$x^{\mu}$) is $V =N (2\pi)^{\tilde{d}/2} \sqrt{\det \theta}$.
Distribution of eigenvalues of $\hat{x}$ is therefore interpreted
as  space-time. In noncommutative space, space-time translation
can be generated by unitary transformation of $A_{\mu}$.
Furthermore,
the dynamical generation of space-time implies that
the fluctuation of space-time is also dynamical and
graviton will be hidden in IIB matrix model or 
equivalently in noncommutative
gauge theory.
Investigations of noncommutative gauge theories have indeed
clarified that they can contain much larger degrees of
freedom than those in the ordinary commutative field theories.
For example,  noncommutative
plane waves  are interpreted as bi-local rather than
local \cite{bilocal}.
After they are expanded
in terms of local operators  it is
expected that higher spin fields will appear, even if we start
from Yang-Mills theory. From this point of view, we expect that noncommutative Yang-Mills
can contain graviton. A possible interpretation of diffeomorphism
invariance in noncommutative Yang-Mills is given in section 3 in
this paper.
\par
In these ways, matrix models can describe both space-time and
matter, i.e. fluctuation around a classical background, in the same
footing. Such a unification seems only possible in the case of
gauge theory where bosonic fields have the same indices as the
space-time. But the unification of space-time and matter is so far 
restricted to a flat space-time
and  a natural question is how we can describe a
curved space-time in matrix models.
(A simple example for the fuzzy sphere is discussed from the matrix
model point of view in \cite{IKTW}.)
In the flat $\tilde{d}$-dimensional space discussed above,
we can identify some of the  $SO(9,1)$ indices with
the indices of  $SO(\tilde{d})$ isometries of the background.
But this cannot be expected for more general curved backgrounds
whose isometries cannot be embedded in $SO(9,1)$. 
If IIB matrix model is a background independent model,
general coordinate transformations will be hidden and
the $SO(9,1)$ symmetries should be rather considered as a gauge fixed
local Lorentz symmetry instead of isometry of 10-dimensional flat space-time.
(A possible interpretation of diffeomorphism in IIB matrix model
is discussed in \cite{ISK}.)
One way to reveal such a structure will be to
find an extended model with larger symmetries that reproduces
IIB matrix model after gauge fixing and integrating irrelevant
fields.  In order that this model can describe  a curved space-time,
a spin connection term containing a $\gamma$-matrix of rank 3
must be included or generated in the fermionic action.

\par
Following the above discussions  we search for models with
higher rank tensor fields  coupled to fermions through 
$\gamma$-matrices.  Another guiding principle to construct a model
is a sufficient
number of supersymmetries.  In order to reproduce IIB matrix model
we need at least 10-dimensional ${\cal N}=2$ supersymmetry.
These requirements can be satisfied by considering matrix models
based on super Lie algebra $osp(1|32,R)$. This superalgebra was
mentioned first on 11-dimensional supergravity\cite{Cremmer:1978km}
and investigated systematically in \cite{vanHolten:1982mx}.
It has attracted a new attention
 as the unified superalgebra for M-theory in \cite{Townsend:1997wg}
\cite{Bergshoeff:2000vg}.
Construction of matrix models based on this superalgebra was
proposed by Smolin \cite{Smolin:2000kc}.
\par
In this paper, we investigate such models, especially from the point
of  supersymmetries. 
In section 2, we consider a model based on $osp(1|32,R)$
super Lie algebra. Bosonic fields in this model can be expanded in terms
of 11-dimensional $\gamma$-matrices of rank 1, 2 and 5. They are real fields.
Fermionic fields are composed of 11-dimensional Majorana fermion.
Hence, reduced to d=10, this model becomes vector-like
and we have to integrate out a right(or left)-handed sector
in order to reproduce IIB matrix model.
The symmetry of this model is a direct product of $OSp(1|32,R)$
and $U(N)$. The $OSp(1|32,R)$ group is a generalization of $SO(9,1)$
in IIB matrix model. The model is also invariant under
constant shifts of fields and we show that the $OSp(1|32,R)$ 
symmetry and the constant shift of fields are combined to form
space-time algebras including space-time supersymmetry.
We discuss  a possibility to obtain IIB matrix model by integrating
some of the fields.
\par
In section 3, we study how diffeomorphism invariance is hidden 
in IIB matrix model. We first give a brief summary of
the relation between matrix models and gauge theories on noncommutative
space. We then show that the unitary gauge transformation is much larger
in noncommutative space than that in ordinary commutative space. 
Even local coordinate transformations are generated
by the unitary transformations.
It is also pointed out that this diffeomorphism invariance is
independent of the global $SO(9,1)$ invariance and it is difficult
to extend the global $SO(9,1)$ to local symmetry of the model.
\par
To search for extended models with local Lorentz invariance,
we then construct matrix models with local $SO(9,1)$ symmetry
in section 4. In particular, we investigate matrix models based
on $u(1|16,16)$ or $gl(1|32,R)$ super Lie algebra. These models are invariant
under coupled symmetries of $U(1|16,16)$ (or $GL(1|32,R)$) and $U(N)$.
Since $U(1|16,16)$ (or $GL(1|32,R)$) is an extension of $SO(9,1)$, this model has
local (i.e. $U(N)$-dependent) Lorentz invariance.
At the cost of this enhanced gauge symmetries, these models break
invariance under constant shifts of fields and we need another
interpretation of space-time translation.
We make use of the Wigner-In\"{o}n\"{u} contraction and identify generators 
of $SO(10,1)$  (which is a subgroup of $U(1|16,16)$ and $GL(1|32,R)$)
with generators of $SO(9,1)$ rotations and translations 
in 10-dimensional space-time. In this way, we can obtain 10-dimensional
space-time picture. We also determine how to scale  the fields 
to obtain the correct 10-dimensional theory.
\par
The final section is devoted to conclusions and discussions.

\section{$osp(1|32,R)$ Cubic Matrix Model}
Smolin proposed a  matrix model based
on the super Lie algebra $osp(1|32,R)$  \cite{Smolin:2000kc}
as an M-theory matrix model. 
% He has investigated a possibility that the model unifies 
% both of superstrings and loop quantum gravity.
The action  is
constructed from $osp(1|32,R)$ matrix  $M$  whose components are also
$N \times N$ matrices. The bosonic part of this
model can be expanded  in terms of 11-dimensional $\gamma$-matrices
with rank 1, 2 and 5.
Therefore it is a natural extension of ordinary matrix models
containing only vector field with rank 1.
Furthermore it has a simple cubic form
in terms of matrix $M$ and is reminiscent of Witten's string field
theory.
Before going into detailed analysis of the model,
we first give the definitions of $osp(1|32,R)$ super Lie algebra.
 \subsection{Definition of $osp(1|32,R)$ supermatrix}
$osp(1|32,R)$ super matrix  is a $33 \times 33$ real supermatrix satisfying
the following conditions:
 \begin{eqnarray}
&& {^{T}M } G + GM = 0 \
  \textrm{  for  }  \ G =  \left( \begin{array}{cc}  \Gamma^{0}  & 0  \\  0
  & i   \end{array} \right), \n
&&M^{\ast} = M.
\label{osp-conditions}
 \end{eqnarray}
$\Gamma^0$ is a $32 \times 32$ 11-dimensional $\gamma$-matrix
in Majorana basis which is real and satisfies $(\Gamma^0)^2=-1$.
%An explicit form is given in appendix. 
For conventions of a super matrix,  see Appendix A.
An element of $OSp(1|32,R)$ group is written as $U= \exp(M)$ and satisfies
\begin{equation}
 {^{T} U} G U = G.
\end{equation}
The above conditions (\ref{osp-conditions}) restrict the matrix $M$ to 
be
\begin{equation}
 M = \left
   ( \begin{array}{cc} m  & \psi \\   i {\bar \psi} & 0 \end{array}
   \right), \label{AZMsu1-16-16}
\end{equation}
where
$\psi$ is a Majorana spinor with 32 components and
${\bar \psi} = {^{T} \psi} \Gamma^{0}$.
$m$ is a real $32 \times 32$ bosonic matrix satisfying
 \begin{eqnarray}
 {^{T}m} \Gamma^{0} + \Gamma^{0} m = 0. \label{AZ41sp32}
 \end{eqnarray}
The bosonic part $m$ is an element of $sp(32,R)$ algebra. It
can be expanded in terms of 11-dimensional $\gamma$-matrices as
     \begin{eqnarray}
     m = u_{A_{1}} \Gamma^{A_{1}} + \frac{1}{2!} u_{A_{1} A_{2}}
      \Gamma^{A_{1} A_{2}} + \frac{1}{5!} u_{A_{1} \cdots
      A_{5} } \Gamma^{A_{1} \cdots A_{5} },
     \label{m-expansion-sp}
     \end{eqnarray}
and contains $528=11+55+462$ degrees of freedom.
$A_i$ are denoted as 11-dimensional indices and run from 0 to 10.
We are working in Majorana basis, where all $\gamma$-matrices are real.
Therefore the real condition means that all the coefficients
$u_{A_{1}}$, $u_{A_{1} A_{2}}$ and $u_{A_{1} \cdots A_{5}}$ are
real.

 \subsection{Action and symmetries}
  In considering the action of a large $N$ reduced model, we regard
  each of the coefficients $u_{A_{1}}$, $u_{A_{1} A_{2}}$ and
  $u_{A_{1} \cdots A_{5}}$ and each component of $\psi$ as an $N
  \times N$ hermitian matrix.  ${M_{P}}^{Q}$, each component of
  the supermatrix $M$, is thus an $N \times N$ hermitian matrix. We
  further introduce 
  $N^{2}$ $osp(1|32,R)$ supermatrices $M^{a}$ as the coefficients of
  $M$ expanded in terms of the Gell-Mann matrices $t^{a}$:
    \begin{eqnarray}
      M = \sum_{a=1}^{N^{2}} t^{a} M^{a}.
    \end{eqnarray}
  The action proposed by Smolin is
  \begin{eqnarray}
   I &=& \frac{i}{g^{2}} Tr_{N \times N} \sum_{Q,R=1}^{33}
  (( \sum_{p=1}^{32} {M_{p}}^{Q} [ {M_{Q}}^{R},
  {M_{R}}^{p}] ) - {M_{33}}^{Q} [ {M_{Q}}^{R}, {M_{R}}^{33} ] )
  \nonumber \\
    &=& \frac{i}{g^{2}} \sum_{a,b,c=1}^{N^{2}} Str_{33 \times 33}
  ( M^{a}  M^{b} M^{c} )  Tr_{N \times N}(t^{a} [t^{b}, t^{c}]),
  \label{AZ42action} 
 \end{eqnarray}
where $p = 1, \cdots, 32$, $P, Q, R= 1, 2, \cdots 33 $ and $a,b,c$
 are indices for $U(N)$. 
 To avoid confusions, we note here that we
use $Tr$ as a trace of $N \times N$  matrices while
 $tr$ (or $Str$) as a trace of $32 \times 32$
($33 \times 33$ super) matrices.
This action can be rewritten as
  \begin{eqnarray}
   I &=& - \frac{f_{abc}}{2g^{2}}  ( tr_{32 \times 32} ( m^{a} m^{b}
   m^{c}) - 3i {\bar  \psi^{a}} m^{b} \psi^{c} )  
\nonumber \\
     &=& \frac{i}{g^{2}} Tr_{N \times N} ( {m_{p}}^{q} [ {m_{q}}^{r} ,
   {m_{r}}^{p} ] ) - 3i {\bar \psi}^p [{m_{p}}^{q}, \psi_{q}] ).   
   \label{AZ42actioncomp} 
  \end{eqnarray}
 $f_{abc}$ are structure constants defined by $[t_a, t_b] = i f_{abc} t_c.$
%Expanding further the bosonic matrix $m$ in terms of
%components  (\ref{m-expansion-sp}),
%we obtain the following action:
%\begin{eqnarray}
%  I &=& - \frac{1}{g^{2}}
%     (112 + 155 + 222 + 255 + 555) + fermion
%\end{eqnarray}
The fermionic term has the same form as that of IIB matrix model but
the bosonic part is cubic and different.
This difference is related to the difference in supersymmetry.
That is,
in the IIB case the supersymmetry transformation (\ref{Ssym1})
for $\psi$ is proportional to
a commutator of the bosonic field $[A_{\mu},A_{\nu}]$
while here all (homogeneous) transformations
are linear in fields as we will see soon. Another big difference is
that this model contains 32 component Majorana fermion compared to
16 in IIB matrix model.
Due to this doubling, we need to integrate out half of fermions
in order to show the equivalence to IIB matrix model.
\par
In spite of these differences,
this model possesses several similarities.
 First it has no free parameter since the coupling constant
is always absorbed by a field redefinition of matrix $M$.
Hence $g$ gives the only dimensionful parameter in the model.
Symmetries of the model also have  similar structures
to IIB matrix model.
If we write the matrix $M$ as a tensor product
of $osp(1|32,R)$ and $N \times N$ matrix,
the action is  invariant under
\begin{equation}
   M  \rightarrow (U \otimes 1_{N \times N}) \ M \
                     (U \otimes 1_{N \times N})^{-1},
\end{equation}
where $U$ is an element of $OSp(1|32,R)$ group.
For an infinitesimal transformation,
\begin{equation}
    \delta M = [H, M]
   = [\left( \begin{array}{cc} h & \chi \\ i {\bar \chi} & 0
   \end{array} \right) ,
      \left( \begin{array}{cc} m & \psi \\ i {\bar \psi} & 0 \end{array}
   \right) ] = \delta_h M + \delta_{\chi}^{(1)} M.
\end{equation}
The bosonic part is identified with $sp(32,R)$ rotations:
\begin{equation}
  \delta_h M = \left( \begin{array}{cc} [h,m] & h\psi \\ i {\bar \psi h}
    & 0 \end{array}
   \right).
\end{equation}
The fermionic part, supersymmetry transformation, is given by
\begin{equation}
   \delta_{\chi}^{(1)} M =
   \left( \begin{array}{cc} i (\chi {\bar \psi} - \psi {\bar
   \chi}) & - m \chi \\ i {\bar \chi} m &  0
\end{array} \right).
\label{hello}
   \end{equation}
The bosonic part is a natural extension
of $SO(9,1)$ rotation in IIB matrix model and indeed includes $SO(10,1)$
symmetry.
 Besides this $SO(10,1)$ symmetry generated by
$\Gamma^{A_1 A_2}$, there are bosonic symmetries
generated by $\gamma$-matrices with rank 1 and 5. These transformations
mix bosonic fields with a different number of 11-dimensional indices.
The fermionic part is a generalization of homogeneous supersymmetry
(\ref{Ssym1}) in IIB matrix model. As already mentioned, this
homogeneous supersymmetry is linear in all fields and 
the action is invariant under this supersymmetry among
terms with the same number of fields.
This is different from IIB matrix model
where the action is balanced between $Tr_{N \times N}[A_{\mu},A_{\nu}]^2$ 
and $Tr_{N \times N}\bar{\psi} \gamma^{\mu}[A_{\mu}, \psi]$ under supersymmetry
since the transformation for the fermion contains two bosonic fields.
We expect that,
 by integrating some of the fields,
the supersymmetry structure of IIB matrix model may be reproduced
from this $osp(1|32,R)$ model.
\par
The action is also invariant under $U(N)$ symmetry
\begin{equation}
   M  \rightarrow (1_{33 \times 33} \otimes U) \ M \
                     (1_{33 \times 33} \otimes U)^{-1},
\end{equation}
where $U$ is an element of $U(N)$ group. All the $osp(1|32,R)$ fields
must be transformed simultaneously.
The symmetry of our model is therefore a direct product
of these two Lie groups $OSp(1|32,R) \times U(N)$.
\par
Another symmetry of the model is a trivial shift of the supermatrix $M$:
\begin{equation}
 {M_{P}}^{Q} \rightarrow  {M_{P}}^{Q}+ {c_{P}}^{Q} 1_{N \times N}.
\end{equation}
This shift contains both bosonic and fermionic inhomogeneous 
transformations.
Some of the bosonic shifts are identified with space-time translations
while the fermionic shifts form space-time supersymmetry
together with the fermionic part of the 
homogeneous $osp(1|32,R)$ transformations.
We write down the fermionic part explicitly for later convenience:
\begin{equation}
 \delta^{(2)}_{\epsilon} m = 0, \ \
        \delta^{(2)}_{\epsilon} \psi = \epsilon.
 \label{fermion-shift}
\end{equation}
\par
Summarizing the three kinds of symmetries,
the bosonic invariance of the model contains
$Sp(32,R)$ rotation with 528 generators,  constant shifts for each
528 $sp(32,R)$ fields and $U(N)$ gauge symmetry.
The fermionic invariance, i.e. supersymmetry, is generated by homogeneous
supersymmetry transformations (\ref{hello}) 
with real 32 components  and
inhomogeneous transformations (\ref{fermion-shift})
with the same number of components.
\par
We then study the  algebraic structures
of these symmetries .
The commutation relations among the homogeneous
supersymmetries (\ref{hello}) are,
of course,  written by $Sp(32,R)$ rotations:
 \begin{eqnarray}
    & &  [ \delta^{(1)}_{\chi}, \delta^{(1)}_{\epsilon} ] m = i
      [(\chi {\bar \epsilon} - \epsilon {\bar \chi}), m],
         \label{AZ43hhm} \\
    & &   [ \delta^{(1)}_{\chi}, \delta^{(1)}_{\epsilon} ]
          \psi = i (\chi{\bar \epsilon} - \epsilon {\bar \chi}) \psi.
      \label{AZ43hhpsi}
 \end{eqnarray}
$h=i (\chi{\bar \epsilon} - \epsilon {\bar \chi})$ is an
element of $sp(32,R)$ and can be expanded as
\begin{equation}
 h =    h_{A} \Gamma^{A} + \frac{1}{2!}
    h_{A_{1} A_{2}} \Gamma^{A_{1} A_{2}}
    + \frac{1}{5!} h_{A_{1} \cdots A_{5}} \Gamma^{A_{1}
    \cdots A_{5}},
\end{equation}
where
   \begin{eqnarray}
       h_{A} = \frac{1}{32} tr(h \Gamma_{A}), \hspace{3mm}
       h_{A_{1} A_{2}} = - \frac{1}{32} tr(h
       \Gamma_{A_{1} A_{2}} ) , \hspace{3mm}
       h_{A_{1} \cdots A_{5}} = \frac{1}{32} tr(h
       \Gamma_{A_{1} \cdots A_{5}} ). \label{AZMAdec7777}
   \end{eqnarray}
It has the same algebraic structure as the
11-dimensional space-time supersymmetry  with central
charges of rank 2 and 5. But this algebra itself can no longer
be interpreted as space-time supersymmetry 
since transformations generated
by $\Gamma^{A}$ are not the translation of space-time.
The situation is the same as in IIB matrix model.
If we interpret eigenvalues of some bosonic variables as our space-time
coordinates, space-time translation should be identified with the
constant shift of bosonic fields.
A difference is that, in IIB matrix model,
this type of commutation relation vanishes up to a field dependent
$U(N)$ gauge transformation while here we have $sp(32,R)$ rotations.
\par
Commutation relations between the homogeneous and inhomogeneous
supersymmetry transformations are, on the other hand,  given by
   \begin{eqnarray}
    [ \delta^{(1)}_{\chi}, \delta^{(2)}_{\epsilon} ] m = -i
    ( \chi {\bar \epsilon} - \epsilon {\bar \chi} ), \hspace{3mm}
    [ \delta^{(1)}_{\chi}, \delta^{(2)}_{\epsilon} ] \psi = 0,
  \label{AZ43SUSYcom69}
   \end{eqnarray}
and generate a constant shift of bosonic fields.
Commutators between inhomogeneous transformations
trivially vanish.
\par
By taking  linear combinations as
\beqa
\tilde{\delta}^{(1)}&=&\delta^{(1)}+\delta^{(2)}, \n
\tilde{\delta}^{(2)}&=&i(\delta^{(1)}-\delta^{(2)}),
\eeqa
we obtain an enhanced 'space-time' supersymmetry algebra
\begin{eqnarray}
[ \tilde{\delta}^{(i)}_{\chi}, \tilde{\delta}^{(j)}_{\epsilon} ] m
   &=&  -2i ( \chi {\bar \epsilon} - \epsilon {\bar \chi})
   \delta_{ij}  ,  \\ \nonumber
[ \tilde{\delta}^{(i)}_{\chi}, \tilde{\delta}^{(j)}_{\epsilon} ] \psi
   &=&  0,
\end{eqnarray}
up to $sp(32,R)$ rotations. As far as the supersymmetry algebra 
is concerned,
$sp(32,R)$ transformations are more appropriately interpreted
as a kind of gauge symmetries.

\subsection{Reduction to $d=10$}
So far we have studied the model from 11-dimensional point of view.
In this subsection, we investigate it from 10-dimensional point of
view by specializing the 10th direction.
For this purpose, we first  introduce the following new
  variables
   \begin{eqnarray}
 & &     W = u_{\sharp}, \hspace{5mm}
    A^{(\pm)}_{\mu} = u_{\mu} \pm  u_{\mu \sharp}, \hspace{5mm}
    C_{\mu_{1} \mu_{2}} = u_{\mu_{1} \mu_{2}},  \\ \nonumber
&&  H_{\mu_{1} \cdots \mu_{4}} = u_{\mu_{1} \cdots \mu_{4} \sharp},
     \hspace{5mm}
    I^{(\pm)}_{\mu_{1} \cdots \mu_{5}} = \frac{1}{2} ( u_{\mu_{1} \cdots
    \mu_{5}} \pm {\tilde u}_{\mu_{1} \cdots \mu_{5}} ).
    \label{AZ43fieldredef}
   \end{eqnarray}
Here we use the indices $\mu_{1}, \mu_{2}, \cdots$ running from
 $0$ to $9$.   $\sharp$ denotes the 10th direction.
The quantity ${\tilde u}_{\mu_{1} \cdots \mu_{5}}$ denotes the
      dual of $u_{\mu_{1} \cdots \mu_{5}}$:
   \begin{eqnarray}
    {\tilde u}_{\mu_{1} \cdots \mu_{5}} = \frac{-1}{5!}
    u_{\mu_{6} \cdots \mu_{10}} \epsilon^{\mu_{1} \cdots \mu_{10} \sharp}.
   \end{eqnarray}
 $I^{(+)}_{\mu_{1} \cdots \mu_{5}}$ and $I^{(-)}_{\mu_{1}
   \cdots \mu_{5}}$ are  self-dual and anti-self-dual respectively:
     \begin{eqnarray}
        I^{(\pm)}_{\mu_{1} \cdots \mu_{5}} = \pm 
     {\tilde I}^{(\pm)}_{\mu_{1} \cdots
        \mu_{5}}.
     \end{eqnarray}
Looking at these fields, we have two set of fields $A_{\mu}^{(\pm)}$ that
can be identified with $A_{\mu}$ field in IIB matrix model.
This is in accord with the doubling of fermions.
These doublings cannot be avoided since we start from a
11-dimensional model.
It is now convenient to define an even rank bosonic field
and two odd rank fields $m_o{}^{(\pm)}$ by
\begin{eqnarray}
 m_e &=& W \Gamma^{\sharp} + \frac{1}{2} C_{\mu \nu} \Gamma^{\mu \nu} 
 + \frac{1}{4!} H_{\mu_1 \cdots \mu_4} \Gamma^{\mu_1 \cdots \mu_4},
\nonumber \\
 m_{o}^{(+)} &=& (\frac{1}{2} A_{\mu}^{(+)} \Gamma^{\mu} + \frac{1}{5!}
 I_{\mu_1 \cdots \mu_5}^{(+)} \Gamma^{\mu_1 \cdots \mu_5})(1+\Gamma^{\sharp}), 
\nonumber \\
 m_{o}^{(-)} &=& (\frac{1}{2} A_{\mu}^{(-)} \Gamma^{\mu} + \frac{1}{5!}
 I_{\mu_1 \cdots \mu_5}^{(-)} \Gamma^{\mu_1 \cdots \mu_5})(1-\Gamma^{\sharp}).
\end{eqnarray}
Fermions are also decomposed into left and right handed chiralities
\begin{equation}
 \psi_L = \frac{1 + \Gamma^{\sharp}}{2} \psi,  \hspace{10mm}
\psi_R = \frac{1 - \Gamma^{\sharp}}{2} \psi.
\end{equation}
Here we note the following useful identities:
\begin{equation}
 m_o^{(+)} \psi_R = m_o^{(-)} \psi_L =0, \ \ 
\bar{\psi}_R m_o^{(+)} = \bar{\psi}_L m_o^{(-)} =0.
\end{equation}
If we denote sets of the fields $m_e, m_o^{(\pm)}$ by ${\cal M}_e, 
{\cal M}_o^{(\pm)}$, we also have the relations
\begin{eqnarray}
&& \chi_L \bar{\epsilon}_L \in {\cal M}_o^{(-)}, \
\chi_R \bar{\epsilon}_R \in {\cal M}_o^{(+)}, \
\chi_L \bar{\epsilon}_R \in {\cal M}_e, \
\chi_R \bar{\epsilon}_L \in {\cal M}_e,
\nonumber \\
&& [{\cal M}_e, {\cal M}_e] \in {\cal M}_e, \
 [{\cal M}_e, {\cal M}_o^{(\pm)}] \in {\cal M}_o^{(\pm)}, \
 [{\cal M}_o^{(+)}, {\cal M}_o^{(-)}] \in {\cal M}_e, \nonumber \\
&& {\cal M}_o^{(+)}  {\cal M}_o^{(+)} =0, \ 
 {\cal M}_o^{(-)}  {\cal M}_o^{(-)} =0.
\end{eqnarray}
Then the action $I = I_{b} + I_{f}$ becomes
\begin{eqnarray}
 I_b &=& \frac{-f_{abc}}{2g^{2}} 
 tr_{32 \times 32}[  m_{e}^a  m_{e}^b m_{e}^c 
  + 3 m_{e}^a  m_{o}^{(+)b}  m_{o}^{(-)c}  
  + 3 m_{e}^a  m_{o}^{(-)b}  m_{o}^{(+)c} ],
\nonumber \\
 I_f &=& \frac{3i f_{abc}}{2 g^{2}}[
 2 {\bar \psi}_{L}^a  m_{e}^b \psi_{R}^c
   +  {\bar \psi}_{L}^a  m^{(+)b}_{o} \psi_{L}^c
    + {\bar \psi}_{R}^a  m^{(-)b}_{o} \psi_{R}^c ].
\label{osp-action-eo}
\end{eqnarray}
The structure is very simple.
There are two sectors ($\psi_{L}$ and $m_o^{(+)}$) and 
($\psi_{R}$ and $m_{o}^{(-)}$) which  are coupled through 
$m_e$ fields. We can then expect to obtain IIB like model 
if we succeed in integrating one sector.
The situation is unfortunately more complicated as we will 
see in the following discussions of supersymmetries. 
Here we write down the action 
in terms of  10-dimensional components for later purpose:
 \begin{eqnarray}
  I_{b} &=& \frac{i}{g^{2}} Tr_{N \times N} ( -96[A^{(+)}_{\mu_{1}} ,
  A^{(-)}_{\mu_{2}} ] C^{\mu_{1} \mu_{2}} 
 - 96 W [A^{(+) \mu} , A^{(-)}_{\mu}] +
  \frac{4}{5} W [ I^{(+)}_{\mu_{1} \cdots \mu_{5}} , I^{(-) \mu_{1} \cdots
  \mu_{5}}]  \nonumber  \\
  &-& 4 H_{\mu_{1} \cdots \mu_{4}} ( [ A^{(+)}_{\mu_{5}} , I^{(-) \mu_{1}
  \cdots \mu_{5}}] - [ A^{(-)}_{\mu_{5}} , I^{(+) \mu_{1} \cdots \mu_{5}}]  )
  - 8 C_{\mu_{1} \mu_{2}} [ {I^{(+)\mu_{1}}}_{\mu_{3} \cdots \mu_{6}} , I^{(-)
  \mu_{2} \cdots \mu_{6}}]  \nonumber \\
  &+&  \frac{8}{3} {H^{\nu \lambda}}_{\mu_{1}
   \mu_{2}} ( [{I^{(+)}}_{\nu\lambda  \mu_{3}
  \mu_{4} \mu_{5}} ,  I^{(-) \mu_{1} \cdots \mu_{5}}] -
  [{I^{(-)}}_{\nu\lambda \mu_{3} \mu_{4} \mu_{5}} ,  I^{(+) \mu_{1} \cdots
  \mu_{5}}]  ) \nonumber \\
  &+& 32 [ {C^{\mu_{1}}}_{\mu_{2}} , C_{\mu_{1} \mu_{3}}] 
 C^{\mu_{2} \mu_{3}} - 16
  C_{\mu_{1} \mu_{2}} [ {H^{\mu_{1}}}_{\mu_{3} \mu_{4} \mu_{5}} , 
 H^{\mu_{2} \cdots
  \mu_{5}}]  \\ \nonumber
  &+& \frac{1}{27} H_{\mu_{1} \cdots \mu_{4}} [ {H^{\nu}}_{\mu_{5} \cdots
  \mu_{7}} , H_{\nu \mu_{8} \mu_{9} \mu_{10}}] \epsilon^{\mu_{1}
  \cdots \mu_{10}  \sharp} ) \nonumber, \\ 
 I_{f} &=& \frac{i}{g^{2}} Tr_{N \times N} (
   -3i (- {\bar \psi_{L}}  [ W , \psi_{R}] +  {\bar \psi_{R}} [ W ,
  \psi_{L}] )  \nonumber \\
  &-& 3i  ({\bar \psi_{L}} \Gamma^{i} [ A^{(+)}_{\mu} , \psi_{L}]
     + {\bar \psi_{R}} \Gamma^{\mu} [ A^{(-)}_{\mu} , \psi_{R}] )
      \nonumber \\
   &-&\frac{3i}{2!} ( {\bar \psi_{L}} \Gamma^{\mu_{1} \mu_{2}}
    [ C_{\mu_{1} \mu_{2}} , \psi_{R}]
  + {\bar \psi_{R}} \Gamma^{\mu_{1} \mu_{2}} [ C_{\mu_{1} \mu_{2}} ,
  \psi_{L}] )   \nonumber \\
 &-& \frac{3i}{4!} ( - {\bar \psi_{L}} 
  \Gamma^{\mu_{1} \mu_{2} \mu_{3} \mu_{4}}
  [ H_{\mu_{1} \mu_{2} \mu_{3} \mu_{4}} ,   \psi_{R}]
   +  {\bar \psi_{R}}
  \Gamma^{\mu_{1} \mu_{2} \mu_{3} \mu_{4}}
 [ H_{\mu_{1} \mu_{2} \mu_{3} \mu_{4}} ,
  \psi_{L}] ) \nonumber \\
   &-& \frac{3i}{5!} ( 2 {\bar \psi_{L}} \Gamma^{\mu_{1} \mu_{2} \mu_{3}
  \mu_{4} \mu_{5}}
 [ {I^{(+)}}_{\mu_{1} \mu_{2} \mu_{3} \mu_{4} \mu_{5}} , \psi_{L}]
    + 2 {\bar \psi_{R}} \Gamma^{\mu_{1} \mu_{2} \mu_{3} \mu_{4} \mu_{5}}
 [ {I^{(-)}}_{\mu_{1} \mu_{2} \mu_{3} \mu_{4} \mu_{5}} ,\psi_{R}] ) ).
  \label{action-10dim}
 \end{eqnarray}
We then investigate the symmetry structures, especially the
structures of supersymmetry, instead of
explicitly integrating out some fields in order to see
a possibility to induce IIB matrix model.
We first perform chiral decomposition of both homogeneous
and inhomogeneous supersymmetries.
64 supersymmetries are decomposed into
16 left(right)-handed homogeneous (inhomogeneous) supersymmetries:
   \begin{eqnarray}
    \delta^{(1)}_{\epsilon_{L.R}}, \ \ \delta^{(2)}_{\epsilon_{L.R}}.
  \end{eqnarray}
Under the homogeneous supersymmetries, the fields transform as
\begin{eqnarray}
   \delta^{(1)}_{\chi} m_e &=& i(\chi_{L}\bar{\psi}_{R} -
   \psi_{L} \bar{\chi}_{R} + \chi_{R}\bar{\psi}_{L} -
   \psi_{R} \bar{\chi}_{L}), 
   \nonumber \\
  \delta^{(1)}_{\chi} m_o^{(+)} &=& i(\chi_{R}\bar{\psi}_{R} -
   \psi_{R} \bar{\chi}_{R}),
 \hspace{10mm}
  \delta^{(1)}_{\chi} \psi_R =  -m_e \chi_R -m_o^{(+)} \chi_L,
 \nonumber \\
  \delta^{(1)}_{\chi} m_o^{(-)} &=&  i(\chi_{L}\bar{\psi}_{L} -
   \psi_{L} \bar{\chi}_{L}),
 \hspace{10mm}
  \delta^{(1)}_{\chi} \psi_L =  -m_e \chi_L - m_o^{(-)} \chi_R.
\label{homosusy-eo}
\end{eqnarray}
Here a natural pairing is
\begin{eqnarray}
  m_{o}^{(+)} \leftrightarrow \psi_{R}, \hspace{10mm}
 m_{o}^{(-)} \leftrightarrow \psi_{L}, 
\label{pairing}
\end{eqnarray}
which is different from the pairing which appears in  the action
(\ref{osp-action-eo}).
The inhomogeneous supersymmetry transformations are trivial
\begin{eqnarray}
   \delta^{(2)}_{\epsilon} \psi_{L(R)} &=& \epsilon_{L(R)},
 \nonumber \\
   \delta^{(2)}_{\epsilon} m &=& 0.
\end{eqnarray}
The commutation relations between the homogeneous supersymmetry
transformations are written in terms of even and odd fields as
   \begin{eqnarray}
    [ \delta^{(1)}_{\chi}, \delta^{(1)}_{\epsilon} ] m_e 
    &=& i [( \chi_L {\bar \epsilon}_R - \epsilon_L {\bar \chi}_R
+ \chi_R {\bar \epsilon}_L - \epsilon_R {\bar \chi}_L   ), m_e]
\nonumber \\
  && + i[ \chi_L {\bar \epsilon}_L - \epsilon_L {\bar \chi}_L , m_o^{(+)}]
  +  i[ \chi_R {\bar \epsilon}_R - \epsilon_R {\bar \chi}_R , m_o^{(-)}],
  \nonumber \\
  \mbox{[} \delta^{(1)}_{\chi}, \delta^{(1)}_{\epsilon}  \mbox{]}  
   m_o^{(+)} 
    &=& i [( \chi_R {\bar \epsilon}_R - \epsilon_R {\bar \chi}_R ), m_e]
 + i( \chi_R {\bar \epsilon}_L - \epsilon_R {\bar \chi}_L ) m_o^{(+)}
-i m_o^{(+)} (\chi_L {\bar \epsilon}_R - \epsilon_L {\bar \chi}_R),
  \nonumber \\
  \mbox{[} \delta^{(1)}_{\chi}, \delta^{(1)}_{\epsilon} \mbox{]}  
  m_o^{(-)} 
    &=& i [( \chi_L {\bar \epsilon}_L - \epsilon_L {\bar \chi}_L ), m_e]
  +  + i( \chi_L {\bar \epsilon}_R - \epsilon_L {\bar \chi}_R ) m_o^{(-)}
-i m_o^{(-)} (\chi_R {\bar \epsilon}_L - \epsilon_R {\bar \chi}_L).
\nonumber \\
\label{homo-eo}
   \end{eqnarray}
The commutation relations between the homogeneous and inhomogeneous
supersymmetry transformations are similarly written as
   \begin{eqnarray}
    [ \delta^{(1)}_{\chi}, \delta^{(2)}_{\epsilon} ] m_e 
    &=& -i ( \chi_L {\bar \epsilon}_R - \epsilon_L {\bar \chi}_R
+ \chi_R {\bar \epsilon}_L - \epsilon_R {\bar \chi}_L   ),
  \nonumber \\
  \mbox{[} \delta^{(1)}_{\chi}, \delta^{(2)}_{\epsilon}  \mbox{]}  
   m_o^{(+)} 
    &=& -i ( \chi_R {\bar \epsilon}_R - \epsilon_R {\bar \chi}_R ),
  \nonumber \\
  \mbox{[} \delta^{(1)}_{\chi}, \delta^{(2)}_{\epsilon} \mbox{]}  
  m_o^{(-)} 
    &=& -i ( \chi_L {\bar \epsilon}_L - \epsilon_L {\bar \chi}_L ).
\label{homo-inhomo-eo}
   \end{eqnarray}
These commutation relations (\ref{homo-inhomo-eo}) 
show that constant shifts of 
the $+(-)$ fields are generated by the right(left)-handed 
supersymmetries. If we neglect the $m_e$ fields, the two sectors
($m_o^{(+)}$ and $\psi_R$) and ($m_o^{(-)}$ and $\psi_L$) 
are completely decoupled.
If we can successfully integrate out $m_e$, $m_o^{(-)}$ and $\psi_L$ 
fields, we expect to obtain a IIB-like matrix model.
As we see from (\ref{homosusy-eo}), if we simply neglect these
fields, the right-handed  homogeneous
supersymmetry transformations for the remaining fields 
($m_o^{(+)}, \psi_R$)  become
\begin{eqnarray}
   \delta^{(1)}_{\chi_R} m_o^{(+)} &=& i(\chi_{R}\bar{\psi}_{R} -
   \psi_{R} \bar{\chi}_{R}),
 \hspace{10mm}
  \delta^{(1)}_{\chi_R} \psi_R =  - \langle m_e \rangle \chi_R.
\end{eqnarray}
Here $\langle m_e \rangle $ should be understood as the vacuum
expectation value expressed in terms of $m_o^{(+)}$.
Hence if $C_{\mu \nu}$ field in $\langle m_e \rangle$ is replaced 
by $[A_{\mu}^{(+)}, A_{\nu}^{(+)}]$, the transformation law can be
identified with the homogeneous supersymmetry in IIB matrix model.

Let us look at these transformations explicitly for 
the rank 1 field $A^{(\pm)}_{\mu}$.
  Under the homogeneous supersymmetry transformation, they transform as
   \begin{eqnarray}
   \delta^{(1)}_{\chi} A^{(+)}_{\mu} &=&
           \frac{i}{8} {\bar \chi_{R}} \Gamma_{\mu} \psi_{R}, \hspace{10mm}
 \delta^{(1)}_{\chi} A^{(-)}_{\mu} =
      \frac{i}{8} {\bar \chi_{L}} \Gamma_{\mu} \psi_{L}.
  \end{eqnarray}
We next consider the commutation relations among supersymmetry
acting on $A^{(\pm)}_{\mu}$ fields.
  Extracting the specific chirality of the supersymmetry parameters, 
   the commutators become
   \begin{eqnarray}
   & & [ \delta^{(1)}_{\chi_{R}}, \delta^{(2)}_{\epsilon_{R}} ]
   A^{(-)}_{\mu} = 0 , \hspace{15mm}
     [ \delta^{(1)}_{\chi_{R}}, \delta^{(2)}_{\epsilon_{R}} ]
   A^{(+)}_{\mu} = \frac{i}{8} {\bar \epsilon}_{R} \Gamma_{\mu} \chi_{R},
 \nonumber \\
  & &  [\delta^{(1)}_{\chi_{L}}, \delta^{(2)}_{\epsilon_{L}} ]
   A^{(-)}_{\mu} = \frac{i}{8} {\bar \epsilon}_{L} \Gamma_{\mu} \chi_{L},
   \hspace{3mm} [\delta^{(1)}_{\chi_{L}}, \delta^{(2)}_{\epsilon_{L}} ]
   A^{(+)}_{\mu} = 0.  \label{AZ43SUSYmainres}
   \end{eqnarray}
This is consistent with the above identification of pairs:
$A^{(-)}_{\mu}$ ($A^{(+)}_{\mu}$) field is paired with the left
(right) chirality. 
  Commutators of  two supersymmetry parameters with different chiralities
   vanish when they act on $A^{(\pm)}_{\mu}$:
   \begin{eqnarray}
       [\delta^{(1)}_{\chi_{L}}, \delta^{(2)}_{\epsilon_{R}} ]
       A^{(\pm)}_{\mu} =  [\delta^{(1)}_{\chi_{R}},
       \delta^{(2)}_{\epsilon_{L}} ]  A^{(\pm)}_{\mu} = 0.
   \end{eqnarray}
We then look at the commutation relations between homogeneous
supersymmetries. We are interested in which generators of
$sp(32,R)$ rotations appear in the commutator.
The commutators between the same chirality
\begin{equation}
          [ \delta^{(1)}_{\chi_{R}}, \delta^{(1)}_{\epsilon_{R}}
          ] A^{(+)}_{\mu}
         =  \frac{i}{8} ( {\bar \chi_{R}}
          [m , \Gamma_{\mu}] \epsilon_{R} )
    \label{AZM43SUSY11a-}
\end{equation}
survive only for  the fields of even
       rank $m_e$ ($W$, $C_{i_{1} i_{2}}$ and $H_{i_{1} \cdots i_{4}}$)
in $m$ in the r.h.s.
Since these fields are integrated out at last, we do not mind
the appearance.
On the other hand, commutators between different chiralities
         \begin{equation}
          [ \delta^{(1)}_{\chi_{L}}, \delta^{(1)}_{\epsilon_{R}}
          ] A^{(+)}_{\mu} =
          \frac{i}{16} ( {\bar \chi_{L}}
          m \Gamma_{\mu} \epsilon_{R} + {\bar \epsilon_{R}}
          \Gamma_{\mu} m  \chi_{L} )  \label{AZM43SUSY11a+}
        \end{equation}
survive for odd rank fields, $A^{(+)}_{\mu}$ and $I^{(+)}_{\mu_{1}
  \cdots \mu_{5}} $, and contains
the field $A^{(+)}_{\mu}$ itself as
  \begin{equation}
   [\delta^{(1)}_{\chi_{L}}, \delta^{(1)}_{\epsilon_{R}}] A^{(+)}_{\mu}
   = \frac{-i}{8} {\bar \chi}_{L}
    A^{(+)}_{\nu} {\Gamma_{\mu}}^{\nu} \epsilon_{R}
   \label{AZ43lrmix+} + \cdots
    \end{equation}
The r.h.s. is generated by $SO(9,1)$ rotation.
Hence, if we interpret the eigenvalue distribution of
$A^{(+)}_{\mu}$ or $A^{(-)}_{\mu}$ as space-time extension,
we need to perform $SO(9,1)$ rotation to obtain the correct
space-time supersymmetry simultaneously with supersymmetries.
In this sense, $SO(9,1)$ symmetry should be more
appropriately considered
as a kind of gauge symmetry. 
%In other words, space-time translation
%and $SO(9,1)$ symmetry is correlated.
\par
The above symmetry arguments support our expectation that
the $osp(1|32,R)$ matrix model becomes IIB matrix model
after integrating out some fields.
However,  there are no terms in the action consistent with
the pairing  expected from the symmetry arguments
(\ref{pairing}). In the next subsection we discuss 
a possibility and also a difficulty to obtain the
correct coupling between fermion and boson
by integrating out unnecessary fields.

\subsection{Integrating out $m_e$, $m_o^{(-)}$ and $\psi_L$ fields}
In order to show that the correct coupling between $\psi_R$
and $A_{\mu}^{(+)}$ can be generated,
we need to integrate out the 
unnecessary fields ($m_e$, $m_o^{(-)}$ and $\psi_L$ ). 
For this purpose, these unnecessary fields
need to have quadratic terms which may be
generated by giving vacuum expectation values
to some fields.
The action (\ref{action-10dim}) indicates that,
if $W$ acquires a vacuum expectation value, 
quadratic terms do not appear for $m_e$ fields.
On the other hand, if $A_{\mu}^{(+)}$ acquires VEV such as
the noncommutative $\hat{p}_{\mu}$, all the unnecessary fields
can get quadratic terms. Hence in the following 
we expand $A^{(+)}_{\mu}$ fields around the noncommutative
 classical background
  \begin{eqnarray}
   A^{(+)}_{\mu} = {\hat p}_{\mu} + a^{(+)}_{\mu},
 \label{classical-background}
  \end{eqnarray}
where ${\hat p}_{\mu}$ satisfy $[{\hat p}_{\mu},{\hat
  p}_{\mu}]=i \beta_{\mu\nu}$ and each $\beta_{\mu\nu}$ is a $c$-number.
  Applying to the action the mapping rule from matrices to functions
(briefly reviewed in the next section), we obtain the following action.
In the following expression, all products are the so-called 
star products and $Tr_{N \times N}$ should be understood as an integral
over noncommutative space.
The quadratic terms are given by
  \begin{eqnarray}
     I_{b}^{(2)} &=& \frac{1}{g^{2}} Tr_{N \times N} ( -96(\partial_{\mu_{1}}
     A^{(-)}_{\mu_{2}} ) C^{\mu_{1} \mu_{2}} + 96 ( \partial_{\mu} W )
     A^{(-) \mu} + 4 (\partial_{\mu_{1}} H_{\mu_{2} \cdots \mu_{5}} ) I^{(-)
     \mu_{1} \cdots \mu_{5}},  \nonumber  \\
I_{f}^{(2)} &=& \frac{1}{g^{2}} Tr_{N \times N} ( - 3i  {\bar
     \psi_{L}} \Gamma^{\mu} \partial_{\mu} \psi_{L} ). \label{2ji}   
 \end{eqnarray}
The cubic interaction terms are given by
\begin{eqnarray}
 I_{b}^{(3)} &=&
  \frac{i}{g^{2}} Tr_{N \times N} ( -96[a^{(+)}_{\mu_{1}} ,
  A^{(-)}_{\mu_{2}} ] C^{\mu_{1} \mu_{2}} - 96 W [a^{(+) \mu} ,
 A^{(-)}_{\mu}] +
  \frac{4}{5} W [ I^{(+)}_{\mu_{1} \cdots \mu_{5}} , I^{(-) \mu_{1} \cdots
  \mu_{5}}] \nonumber \\
  &+& 4 ( [ a^{(+)}_{\mu_{1}} , H_{\mu_{2} \cdots \mu_{5}}] I^{(-) \mu_{1}
  \cdots \mu_{5}} - [ A^{(-)}_{\mu_{1}} , H_{\mu_{2} \cdots \mu_{5}}] I^{(+)
  \mu_{1} \cdots \mu_{5}}  )
    - 8 C_{\mu_{1} \mu_{2}} [ {I^{(+)\mu_{1}}}_{\mu_{3} \cdots \mu_{6}} ,
 I^{(-)
  \mu_{2} \cdots \mu_{6}}]  \nonumber \\
  &+&  \frac{8}{3} {H^{\nu \lambda}}_{\mu_{1} \mu_{2}} 
( [{I^{(+)}}_{\nu \lambda \mu_{3}
     \mu_{4} \mu_{5}} ,  I^{(-) \mu_{1} \cdots \mu_{5}}] -
  [{I^{(-)}}_{\nu \lambda  \mu_{3} \mu_{4} \mu_{5}} ,  I^{(+) \mu_{1} \cdots
  \mu_{5}}]  ) \nonumber \\
  &+& 32 [ {C^{\mu_{1}}}_{\mu_{2}} , C_{\mu_{1} \mu_{3}}] 
C^{\mu_{2} \mu_{3}} - 16
  C_{\mu_{1} \mu_{2}} [ {H^{\mu_{1}}}_{\mu_{3} \mu_{4} \mu_{5}} , 
H^{\mu_{2} \cdots
  \mu_{5}}]
\nonumber \\
  &+& \frac{1}{27} H_{\mu_{1} \cdots \mu_{4}} [ {H^{\nu}}_{\mu_{5} \cdots
  \mu_{7}} , H_{\nu \mu_{8} \mu_{9} \mu_{10}}] \epsilon^{\mu_{1}
 \cdots \mu_{10} \sharp} ), \label{AZ44others} \\
  I_{f}^{(3)}  
  &=& \frac{i}{g^{2}} Tr_{N \times N} (-3i ( - {\bar \psi_{L}} [ W ,
     \psi_{R}] +  {\bar  \psi_{R}} [ W , \psi_{L}] )
   - 3i ( {\bar \psi_{L}} \Gamma^{\mu} [ a^{(+)}_{\mu} , \psi_{L}]
     + {\bar \psi_{R}} \Gamma^{\mu} [ A^{(-)}_{\mu} , \psi_{R}] )
 \nonumber \\
   &-&\frac{3i}{2!} ( {\bar \psi_{L}} \Gamma^{\mu_{1} \mu_{2}}
 [ C_{\mu_{1} \mu_{2}} ,
  \psi_{R}] + {\bar \psi_{R}} \Gamma^{\mu_{1} \mu_{2}} [ C_{\mu_{1} \mu_{2}} ,
  \psi_{L}] )
\nonumber \\
   &-& 
\frac{3i}{4!} ( - {\bar \psi_{L}} \Gamma^{\mu_{1} \mu_{2} \mu_{3} \mu_{4}}
  [ H_{\mu_{1} \mu_{2} \mu_{3} \mu_{4}} ,   \psi_{R}] +  {\bar \psi_{R}}
  \Gamma^{\mu_{1} \mu_{2} \mu_{3} \mu_{4}}
 [ H_{\mu_{1} \mu_{2} \mu_{3} \mu_{4}} ,
  \psi_{L}] ) \nonumber \\
   &-& \frac{3i}{5!} ( 2 {\bar \psi_{L}} \Gamma^{\mu_{1} \mu_{2} \mu_{3}
     \mu_{4} \mu_{5}}
 [ {I^{(+)}}_{\mu_{1} \mu_{2} \mu_{3} \mu_{4} \mu_{5}} , \psi_{L}]
   +  2 {\bar \psi_{R}} \Gamma^{\mu_{1} \mu_{2} \mu_{3} \mu_{4} \mu_{5}}
 [ {I^{(-)}}_{\mu_{1} \mu_{2} \mu_{3} \mu_{4} \mu_{5}} ,\psi_{R}] ) ).
     \label{AZ44fermother}
 \end{eqnarray} 
 \begin{figure}[htbp]
   \begin{center}
    \scalebox{.6}{\includegraphics{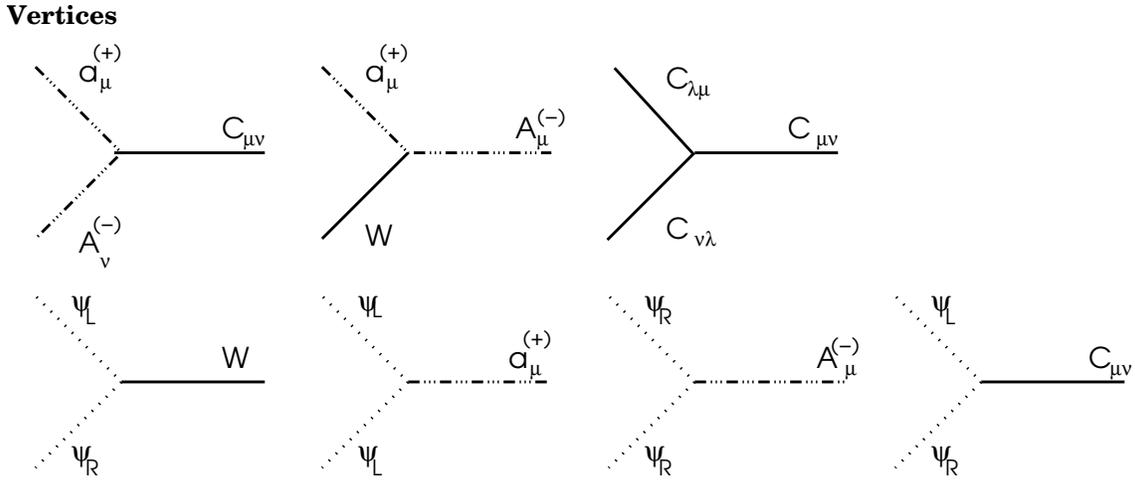} }
   \end{center}
    \caption{Typical vertices of the $osp(1|32,R)$ matrix model}
   \label{feynvertex}
  \end{figure}
We draw some typical vertices in fig. \ref{feynvertex}.
\par
\begin{figure}[htbp]
   \begin{center}
    \scalebox{.6}{\includegraphics{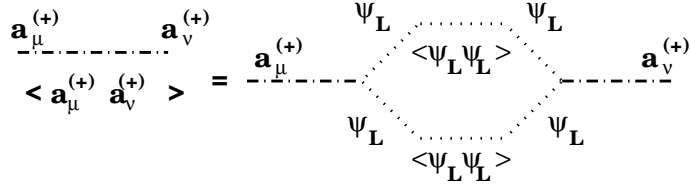} }
   \end{center}
    \caption{A propagator $\langle a^{(+)}_{\mu} a^{(+)}_{\nu}
      \rangle$ is induced by one-loop effect.}
   \label{propwwaa}
  \end{figure}

If we neglect the interaction terms first, an integration over $C_{\mu \nu}$
field gives a constraint on $A^{(-)}$ as
\begin{equation}
 \partial_{\mu} A_{\nu}^{(-)} -  \partial_{\nu} A_{\mu}^{(-)} =0,
\end{equation}
and we can solve it in terms of a scalar 
as $A_{\mu}^{(-)}= \partial_{\mu} \lambda$.
Inserting this into the quadratic term, we have a propagator connecting
two scalars, $\lambda$ and $W$. 
Then we can integrate unnecessary fields $A_{\mu}^{(-)}, W$ 
$I^{(-)}, H$ and $\psi_L$. 
An issue here is whether we can generate  the IIB-like terms
such as $Tr_{N \times N} [A_{\mu}^{(+)}, A_{\nu}^{(+)}]^2$ or 
$Tr_{N \times N} \bar{\psi}_R \Gamma^{\mu} [A_{\mu}^{(+)},\psi_R]$.
$U(N)$ gauge symmetry assures the existence of these terms
 if we can show that 
 quadratic kinetic terms for $a_{\mu}^{(+)}$ and $\psi_R$, that is,
$(\partial_{\mu} a_{\nu}^{(+)} - \partial_{\nu} a_{\mu}^{(+)})^2$
and $\bar{\psi}_R \Gamma^{\mu} \partial_{\mu} \psi_R$, are generated.
First, the kinetic term for $a_{\mu}^{(+)}$ is easily generated by
integrating out $\psi_L$ as in fig. \ref{propwwaa}, since there is a vertex
$\bar{\psi}_L \Gamma^{\mu} [a_{\mu}^{(+)}, \psi_L]$.\\

\begin{figure}[htbp]
   \begin{center}
    \scalebox{.6}{\includegraphics{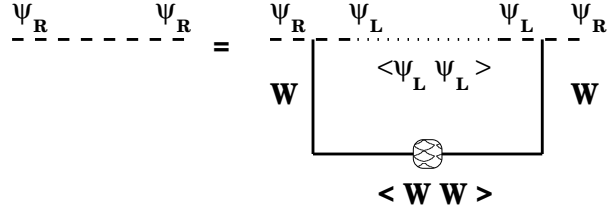} }
   \end{center}
    \caption{A propagator for $\psi_{R}$ is induced if $W$ can acquire 
      a propagator.}
   \label{truefinalanswer}
  \end{figure}

The kinetic term for $\psi_R$ is more difficult to generate.
As is seen from fig. \ref{truefinalanswer},
 one way to generate such a
term is to connect two $\bar{\psi}_L [W, \psi_R]$ vertices 
by propagators of $\psi_L$ and $W$. \\

Propagators for $W$ and $\lambda$ fields cannot be generated
perturbatively as we prove in Appendix B.
But since there is no symmetry prohibiting such terms
it does not exclude a nonperturbative generation.
The existence of the propagator connecting these two scalar fields
rather
indicate that both of these two propagators can be generated
self-consistently.
We do not discuss more details here, but it is 
probable that $W$ acquires a propagator and the above mentioned 
kinetic term for $\psi_R$ will be also generated.
In this way, we expect that IIB matrix model is induced from
$osp(1|32,R)$ model.

\section{Diffeomorphism in noncommutative Yang-Mills}

In this section, we first review how noncommutative Yang-Mills
is obtained from matrix models and then investigate the special properties
regarding the local gauge transformations in noncommutative background.
Especially we study special types of gauge transformations
which can be interpreted as local coordinate transformations.
\par
First we give a brief review on a matrix model description of
noncommutative field theories.
The noncommutative background $\hat{x}$ satisfying
\begin{equation}
 [\hat{x}^{\mu},  \hat{x}^{\nu}] = - i \theta^{\mu \nu}
\label{noncommutative-x2}
\end{equation}
with a c-number $\theta^{\mu\nu}$  
is a classical solution of IIB matrix model.
We assume the rank of $\theta^{\mu\nu}$
to be $\tilde{d}$ and define its inverse $\beta_{\mu\nu}$ in
$\tilde{d}$-dimensional  subspace.
This expression is  formal and only valid for infinite $N$.
For finite $N$, see papers \cite{nishi-nc}.
$\hat{x}^{\mu}$ satisfy the canonical commutation relations and
they span the $\tilde{d}$-dimensional phase space.
Therefore the momentum operators are proportional to the coordinates as
\begin{equation}
 \hat{p}_{\mu} = \beta_{\mu \nu} \hat{x}^{\nu}.
\label{noncommutative-p}
\end{equation}
The semiclassical correspondence shows that the
volume of the phase space (measured in the coordinate space of
$x^{\mu}$) is $V =N (2\pi)^{\tilde{d}/2} \sqrt{ \det \theta}$.
We  expand the bosonic matrices
$A^{\mu}$ around $\hat{x}^{\mu}=\theta^{\mu \nu} \hat{p}_{\nu}$ as 
\begin{equation}
   A^{\mu} = \theta^{\mu \nu} ( \hat{p}_{\nu} + \tilde{a}_{\nu}).
\label{a-expand-flat}
\end{equation}
If we assume that all fields can be expanded in terms
of noncommutative plane wave $\exp(ik \cdot \hat{x})$,
we obtain a map from a matrix
\begin{equation}
 \hat{a} = \sum_k  \tilde{a}(k) \exp( i k \cdot \hat{x})
\label{operator}
\end{equation}
to a function
\beq
 a(x)=\sum_k \tilde{a}(k) \exp( i k \cdot x)
\label{proj}
\eeq
in the $\tilde{d}$-dimensional noncommutative plane.
By this construction, a product of matrices is mapped to the $\star$
product of functions
\beqa
\hat{a}\hat{b} &\rightarrow& a(x)\star b(x),\n
a(x) \star b(x) &\equiv& \exp({\theta^{\mu \nu}\over 2i}{\partial ^2\over
\partial\xi^{\mu}
\partial\eta^{\nu}})
a(x+\xi )b(x+\eta )|_{\xi=\eta=0}
\label{star}
\eeqa
and the operation $Tr$ over matrices can be exactly mapped onto the integration
over functions as
\beq
Tr[\hat{a}] =
\sqrt{\det \beta}({1\over 2\pi})^{\tilde{d}\over 2}\int d^{\tilde{d}}x \
a(x) .
\label{traceint}
\eeq

The reduced model can be shown to be equivalent to
noncommutative Yang-Mills by 
the following map from matrices onto functions
\beqa
\hat{a} &\rightarrow& a(x) , \n
\hat{a}\hat{b}&\rightarrow& a(x)\star b(x) ,\n
Tr&\rightarrow&
\sqrt{\det \beta}({1\over 2\pi})^{\tilde{d}\over 2}\int d^{\tilde{d}}x .
\label{momrule}
\eeqa
Applying the rule eq.(\ref{momrule}),
we can obtain  $U(1)$ gauge theory on
$\tilde{d}$-dimensional noncommutative space.
(Noncommutative $U(m)$ gauge theory can be similarly obtained
by expanding around $x^{\mu} \otimes 1_{m}$.)
\par
The extension of $\hat{x}^{\mu}$ can be interpreted as the space-time and
the space-time translation is realized by the following unitary
operator:
\beq
\exp(i\hat{p}\cdot \epsilon)\hat{x}^{\mu}
\exp(-i\hat{p}\cdot \epsilon)\n
=\hat{x}^{\mu}+\epsilon^{\mu} .
\label{transl}
\eeq
It is amusing  that the translation in the noncommutative space is realized
by $U(N)$ gauge transformations in matrix models. This realization
 has been known as Parisi prescription in the old reduced models
 \cite{parisi}  and reinvestigated
 \cite{rey}\cite{Gross:2000ba} from the noncommutative point of view
to study extended gauge invariant operators (open Wilson lines).
The local gauge symmetry of noncommutative Yang-Mills is originated in
the invariance under $U(N)$ invariance of IIB matrix model
\begin{equation}
 A_{\mu}\rightarrow UA_{\mu}U^{\dagger}.
\end{equation}
Indeed, if we expand $U= \exp (i\hat{\lambda} )$ and parameterize
${\hat \lambda}$ as 
\begin{equation}
\hat{\lambda}=\sum_k \tilde{\lambda} (k)
 \exp( i{k} \cdot\hat{x}),
\label{gauge-parameter}
\end{equation}
we find that the fluctuating field of $A_{\mu}$ around
the fixed noncommutative background transforms as
\begin{equation}
\hat{a}_{\mu} \rightarrow  \hat{a}_{\mu}+i[\hat{p}_{\mu},\hat{\lambda} ]
-i[\hat{a}_{\mu},\hat{\lambda} ].
\end{equation}
After mapping   the transformation onto functions, we have
\begin{eqnarray}
&& a_{\alpha}(x) \rightarrow a_{\alpha}(x) +
{\partial \over \partial x^{\alpha}}\lambda (x)
-i[a_{\alpha}(x), \lambda (x)]_{\star}, \n
&& a_{i} \rightarrow a_i -i[a_{i}(x), \lambda (x)]_{\star}, \n
&& \psi \rightarrow \psi  -i[\psi(x), \lambda (x)]_{\star},
\label{gauge-tr}
\end{eqnarray}
where $1 \le \alpha \le \tilde{d}$ and $i > \tilde{d}$.
If we take $\lambda$ as in (\ref{transl}),
\begin{equation}
 \hat{\lambda} = \epsilon^{\alpha} \hat{p}_{\alpha},
\label{global-shift}
\end{equation}
the transformations
(\ref{gauge-tr}) become translation in the noncommutative space
up to a constant shift of the gauge field:
\begin{eqnarray}
&&  a_{\alpha}(x) \rightarrow a_{\alpha}(x)
- \beta_{\alpha \beta} \epsilon^{\beta}
+ \epsilon^{\beta} \partial_{\beta} a_{\alpha}(x), \n
&&  a_i(x) \rightarrow a_i(x) +   \epsilon^{\beta} \partial_{\beta} a_i(x), \n
&& \psi(x) \rightarrow \psi(x) +   \epsilon^{\beta} \partial_{\beta} \psi(x).
\end{eqnarray}
As the above example shows, local gauge symmetries in noncommutative
gauge theories are very different from those in the ordinary gauge
theories and
even space-time translation is generated.
Hence 
all gauge invariant operators are invariant under the space-time
translation and 
should be constructed by integrating over space-time, which
is reminiscent of the theory of gravity.
We are, therefore, tempted to generalize
the above discussion to local transformations.
The reason why the gauge transformations in noncommutative space-time
are much larger than those in commutative space is that gauge
transformation parameters $\lambda$ contain not only ordinary functions but
differential operators in the semiclassical limit in the following sense.
 Functions in noncommutative space are expanded
in terms of  noncommutative plane waves as in (\ref{gauge-parameter}).
For finite $N$, the momenta of plane waves take values of
 \begin{equation}
  k_n = \sqrt{2\pi \beta} N^{-1/\tilde{d}} \ n,
\label{momentum-region}
 \end{equation}
where $n =0,1, \cdots, N^{2/\tilde{d}}$. This can be seen from the
explicit construction of the plane waves in terms of 't Hooft matrices
$U$ and $V$ satisfying
$UV=\exp(2\pi i /N) VU$.
The number of independent plane waves is $N^2$, which is
the same as the number of degrees of freedom of a hermitian matrix.
Since the natural cut off scale in the noncommutative plane
(\ref{noncommutative-x}) is given by
$l_0=\sqrt{2 \pi \theta} = \sqrt{2\pi / \beta}$,
the natural cut off of momenta should be $2 \pi/l_0=\sqrt{2\pi \beta}$.
However, some plane waves with momenta $(\ref{momentum-region})$
exceed this natural bound and they become very nonlocal objects
since such high momentum plane waves generate translation in
space with $l_0 N^{-1/\tilde{d}} n$.
Hence only $N$ out of $N^2$ plane waves
whose momenta are smaller than $\sqrt{2 \pi \beta}$ 
can be interpreted as ordinary
plane waves in the semiclassical limit and others should be
interpreted as differential operators that can generate
translation in noncommutative direction in space-time.
From a matrix model point of view, such nonlocal waves correspond
to off-diagonal elements while local ones to diagonal elements.
\par
Now let us consider the gauge transformations related to local
coordinate transformations.
A natural generalization of the global translation (\ref{global-shift})
will be
\begin{equation}
 \lambda =  \frac{1}{2}(\hat{p}_{\alpha}
    \hat{\epsilon}^{\alpha}
 + \hat{\epsilon}^{\alpha}  \hat{p}_{\alpha} ),
\label{local-translation}
\end{equation}
or, if we want to include both of the gauge transformations and
local coordinate transformations in the semiclassical limit,
we can expand $\lambda$ as
\begin{equation}
 \lambda = \hat{\lambda}_0 + \frac{1}{2}(\hat{p}_{\alpha}
    \hat{\epsilon}^{\alpha}
 + \hat{\epsilon}^{\alpha}  \hat{p}_{\alpha} ).
\end{equation}
Similarly we expand bosonic matrices as
\begin{equation}
 \tilde{A}_{\mu} =\beta_{\mu \nu} A^{\nu} = \tilde{a}_{\mu} + \frac{1}{2}(\hat{p}_{\alpha}
    \hat{e}_{\mu}^{\alpha}
 + \hat{e}_{\mu}^{\alpha}  \hat{p}_{\alpha} ),
\end{equation}
and assume that the field $\hat{e}_{\mu}^{\alpha}$ is close to
$\delta_{\mu}^{\alpha}$:
\begin{equation}
 \hat{e}_{\mu}^{\alpha} = \delta_{\mu}^{\alpha} + h_{\mu}^{\alpha}.
\end{equation}
This is a natural generalization of the expansion
(\ref{a-expand-flat}).
Applying the unitary transformation generated by (\ref{local-translation})
to the bosonic field expanded as above, we have the following
transformation law:
\begin{eqnarray}
 \delta h_{\mu}^{\alpha}(x) &=&
   - e_{\mu}^{\beta} (x) \partial_{\beta} \epsilon^{\alpha}
   + \epsilon^{\beta} \partial_{\beta} h_{\mu}^{\alpha}, \n
 \delta a_{\mu}(x) &=&   \epsilon^{\beta} \partial_{\beta} a_{\mu}(x).
\label{diffeo}
\end{eqnarray}
Here we have assumed that all of $\epsilon^{\alpha}, a_{\mu}$ 
and $e^{\alpha}_{\mu}$ are slowly varying and dropped higher
derivative terms.
In addition to the transformations of the fields, we need to
transform the background as
\begin{equation}
 \delta \hat{p}_{\alpha} = \beta_{\alpha \beta} \hat{\epsilon}^{\beta}
\end{equation}
or $\delta \hat{x}^{\mu}=\epsilon^{\mu}$ in terms of $\hat{x}^{\mu}$.
Accordingly, the commutation relations between $\hat{p}_{\mu}$
change as
\begin{equation}
 [\hat{p}_{\alpha}+  \beta_{\alpha \beta} \hat{\epsilon}^{\beta},
 \hat{p}_{\alpha'}+ \beta_{\alpha' \beta'} \hat{\epsilon}^{\beta'} ]
 = i (\beta_{\alpha \alpha'}
    + \beta_{\beta \alpha'} \partial_{\alpha} \epsilon^{\beta}
      + \beta_{ \alpha \beta} \partial_{\alpha'} \epsilon^{\beta} ).
\end{equation}
Therefore the shift of the background
 can be interpreted as the transformation of
$\beta_{\alpha \alpha'}$.
\par
The transformation (\ref{diffeo}) indicates that local coordinate
transformations can be realized by gauge transformations in
noncommutative space. But as the transformations (\ref{diffeo}) 
show, the gauge field and the
fermion field transform as a scalar and ${h_{\mu}}^{\alpha}$
as a vector field. In other words, the $SO(9,1)$ index has
nothing to do with this coordinate transformation as it should be
since $SU(N)$ transformations and $SO(9,1)$ transformations are
independent from the beginning. IIB matrix model is not explicitly
invariant under local $SO(9,1)$. 
Local Lorentz transformations might be realized in a
complicated way in IIB matrix model and
we expect that there is an extended model with
obvious local Lorentz symmetry, 
which becomes IIB matrix model after gauge fixing.
In the next section, we search for such models.

\section{Gauged matrix models}
In this section we investigate another type of matrix models with
larger {\it local} symmetries. The model studied in section 2 has an extension
of $SO(9,1)$ symmetry, that is, $OSp(1|32,R)$ symmetry. But this symmetry is
decoupled from $U(N)$ gauge symmetry. As we have seen in section 3,
space-time is realized as eigenvalues of bosonic matrices and consequently
some $U(N)$ symmetry is identified with space-time  translation.
Hence if local Lorentz symmetry exists it should be $U(N)$ dependent
$SO(9,1)$ symmetry and we need to unify decoupled $SO(9,1)$ and
$U(N)$ invariance in IIB matrix model.
\par
Let us first try to
gauge the global $SO(9,1)$ symmetry. A convenient way to write $SO(9,1)$
is to use  $\gamma$-matrices. Defining
\begin{equation}
 m= \Gamma^{\mu} A_{\mu}, \ \
 h=\frac{1}{2} \zeta_{\mu\nu} \Gamma^{\mu\nu},
\label{gauging-start}
\end{equation}
$SO(9,1)$ rotations of $A_{\mu}$ and $\psi$ are given by
\begin{equation}
\delta m = [h, m], \ \ \delta \psi = h \psi, \ \ \delta \bar{\psi}
 = - \bar{\psi} h.
\label{so10}
\end{equation}
The rotation angle $\zeta_{\mu\nu}$ is a $c$-number.
% Gauging $SO(9,1)$ causes some problems in matrix models.
One  way to gauge global symmetries in matrix models is
to make transformation parameters $U(N)$ dependent.
A big difference here from local gauge symmetries in ordinary
commutative space-time is that $U(N)$-dependent matrices
are generally not commutative while $x$-dependent local
parameters are of course commutative.
Therefore, the algebra does not close within the original
transformations. In our case of $SO(9,1)$, since
\begin{equation}
  [\hat{h},\hat{h'}] =
[\frac{1}{2}\hat{\zeta}_{\mu\nu} \Gamma^{\mu\nu},
 \frac{1}{2}\hat{\zeta}_{\mu'\nu'} \Gamma^{\mu'\nu'}]
=\frac{1}{8} ([\Gamma^{\mu\nu},\Gamma^{\mu'\nu'}]
\{\hat{\zeta}_{\mu\nu},\hat{\zeta}_{\mu'\nu'}\}
+\{\Gamma^{\mu\nu},\Gamma^{\mu'\nu'}\}
[\hat{\zeta}_{\mu\nu},\hat{\zeta}_{\mu'\nu'}]),
\end{equation}
and the commutator between $\hat{\zeta}_{\mu\nu}$ does not
vanish, we need to include transformations generated by the
anti-commutators of $\Gamma^{\mu\nu}$, that is, $1$ and
$\Gamma^{\mu_1 \mu_2 \mu_3 \mu_4}$. 
Repeating this procedure,
the algebra finally closes in the $\gamma$-matrices with even rank, 
$1$, $\Gamma^{\sharp}$, $\Gamma^{\mu\nu}$, $\Gamma^{\mu \nu \sharp}$,
$\Gamma^{\mu_1 \mu_2 \mu_3 \mu_4}$ and $\Gamma^{\mu_{1} \mu_{2}
  \mu_{3} \mu_{4} \sharp}$.
There are 512 bosonic generators.
The coefficients $\zeta_{\mu \nu}$ must be extended to complex
matrices. 
Since these transformations can be restricted to chiral sectors of
fermions,
we can obtain closed gauged algebra acting on Weyl fermions
of IIB type. Generalizing this bosonic algebra by including
supersymmetries, we obtain $gl(1|16,C)$ super Lie algebra.
As far as the algebras are concerned, $gl(1|16,C)$ is a minimal 
gauged extension of $so(9,1)$.
As for dynamical fields, if we start from a vector boson
with rank 1 $\gamma$-matrix, $gl(16,C)$ bosonic transformations
generate fields with other odd rank $\gamma$-matrices and
we have to include $A_{\mu_1 \mu_2 \mu_3}$ and 
$A_{\mu_1 \cdots \mu_5}$ in addition to $A_{\mu}$.
There are 256 bosonic fields.
A model based on this $gl(1|16,C)$ super Lie algebra is an interesting
possibility, but it turns out difficult to find an invariant action.
In the following we instead investigate a model with 
local $osp(1|32,R)$  gauge symmetry. 
That is, we demand that the model should be invariant under
$U(N)$ dependent $osp(1|32,R)$ symmetry.
This model has larger symmetries and more fields than the $gl(16,C)$
model, but the invariant action can be easily constructed in terms of 
supermatrices as we show below.
We must extend the chiral fermions to
include both chiralities. 
We also have to extend the bosonic degrees of freedom 
$m$ by including fields with all ranks in 11 dimensions.
\par
In constructing an invariant action of the gauged matrix model,
it is generally difficult to keep both the gauge symmetries and 
invariance under a constant shift of fields.
If the fields transform as in (\ref{so10})
after gauging, the action of the type
$Tr_{N \times N}( \bar\psi m \psi)$ or $Tr_{N \times N}(m^3)$
are invariant.
However, the action such as
$Tr_{N \times N}(\bar\psi \Gamma^{\mu} [A_{\mu},\psi])$ is not invariant
under gauge transformations
and it is difficult to keep  both invariances. 
In this paper we abandon the latter invariance and
consider the action
   \begin{eqnarray}
        I = \frac{1}{g^{2}} Tr_{N \times N} Str_{33 \times 33}
        ( M^{3} ). \label{AZ51action}
   \end{eqnarray}
This action was also proposed by Smolin. 
We call it a gauged model  because
  it is invariant under local $osp(1|32,R)$ symmetry, that is,
    a tensor product of two gauge symmetries  $osp(1|32,R)$ and $u(N)$.
   Instead of this  enhancement of the symmetries, this action
   is not invariant under a constant shift of field.
   This looks troubling since, as we saw in
   section 2,  commutators between the homogeneous and the inhomogeneous
  supersymmetries generate  a space-time translation, a constant shift 
   of bosonic field, and if we lose inhomogeneous translational invariance
   of bosons and fermions
   we may also lose space-time interpretation of supersymmetries.
   However, this problem can be resolved by identifying some generators
   of $osp(1|32,R)$ (or its extension $u(1|16,16)$ ) with space-time
   translation generators using the Wigner-In{\"o}n{\"u} contraction.

   There are two ways to gauge the $osp(1|32,R)$ model.
   One way proposed by Smolin is to use $u(1|16,16)$ super Lie algebra,
  a complexification of $osp(1|32,R)$.
  He conjectured that this gauged model describes loop quantum gravity
  \cite{Smolin:2000kc}. 
  Another way to gauge is to use $gl(1|32,R)$ super Lie algebra,
  an analytic continuation  of  $u(1|16,16)$.
We next see the definitions of these super Lie algebras and also see
why they are gauged symmetries of the global $osp(1|32,R)$.

 \subsection{Definitions of $u(1|16,16)$ and $gl(1|32,R)$}
 An element  $M$  of  $u(1|16,16)$ super Lie algebra satisfies
 \begin{eqnarray}
   M^{\dagger} G + GM = 0 \
  \textrm{ for } \ G =  \left( \begin{array}{cc}  \Gamma^{0}  & 0  \\  0
  & i   \end{array} \right).
 \end{eqnarray}
The reality condition is not imposed in this case.
 The above definition restricts the $33 \times 33$ matrix form of $M$ as
  \begin{eqnarray}
   M = \left
   ( \begin{array}{cc} m  & \psi \\   i {\bar \psi} & v \end{array}
   \right), \label{AZ51su11616}
  \end{eqnarray}
  where $v$ is pure imaginary, $\psi$ is a general complex spinor 
  and ${\bar \psi} = \psi^{\dagger} \Gamma^{0}$.
  The bosonic part $m$ can be expanded in terms of 11-dimensional $\gamma$-matrices,
     \begin{eqnarray}
         m &=& u {\bf 1} + u_{A_{1}} \Gamma^{A_{1}} + \frac{1}{2!}
         u_{A_{1} A_{2}} \Gamma^{A_{1} A_{2}} + \frac{1}{3!}
         u_{A_{1} A_{2} A_{3}} \Gamma^{A_{1} A_{2} A_{3}} \nonumber \\
         &+& \frac{1}{4!} u_{A_{1} \cdots A_{4}} \Gamma^{A_{1}
         \cdots A_{4}} + \frac{1}{5!} u_{A_{1} \cdots A_{5}}
         \Gamma^{A_{1} \cdots A_{5}}, 
   \label{m-expansion-11}
        \end{eqnarray}
where  $u_{A_{1}}$, $u_{A_{1} A_{2}}$ and
      $u_{A_{1} \cdots A_{5}}$ are real, while $u$, $u_{A_{1}
      A_{2} A_{3}}$ and $u_{A_{1} \cdots A_{4}}$ are
      pure imaginary.
Pure imaginary valued coefficients $u$, $u_{A_{1}
      A_{2} A_{3}}$ and $u_{A_{1} \cdots A_{4}}$ 
     are new compared to $osp(1|32,R)$.
 Fermions are also doubled since we do not impose the Majorana condition.
  This matrix can be decomposed into two  matrices
\begin{eqnarray}
  M = H + A', \ \
    H =
   \left( \begin{array}{cc} m_{h} & \psi_{h} \\ i {\bar \psi_{h}} & 0
   \end{array} \right), \hspace{3mm}
 A'=
    \left( \begin{array}{cc} m_{a} & i \psi_{a} \\ {\bar \psi_{a}}
   & iv \end{array} \right),
\end{eqnarray}
where
   \begin{eqnarray}
 m_{h} &=& u_{A_{1}} \Gamma^{A_{1}} +
   \frac{1}{2!} u_{A_{1} A_{2}} \Gamma^{A_{1} A_{2}} +
   \frac{1}{5!} u_{A_{1} \cdots A_{5}} \Gamma^{A_{1} \cdots
   A_{5}}, 
   \nonumber \\
m_{a} &=& u + \frac{1}{3!} u_{A_{1}
   A_{2} A_{3}} \Gamma^{A_{1} A_{2} A_{3}} + \frac{1}{4!}
   u_{A_{1} \cdots A_{4}} \Gamma^{A_{1} \cdots
   A_{4}},  
\label{mhma}
   \end{eqnarray}
and $\psi_{h}$ and $\psi_a$ are real fermions.
They satisfy the following relations
   \begin{eqnarray}
     {^{T} H} G + G H = 0,  \hspace{3mm}
     {^{T} A}' G - G A' = 0.
     \label{AZ51hasu}
   \end{eqnarray}
  The matrix  $ H$ forms  $osp(1|32,R)$ super Lie subalgebra
  of $u(1|16,16)$ algebra but $A'$ does not form an algebra by themselves.
  We denote the former set of matrices by
  ${\cal H}$ and the latter by ${\cal A'}$.
  Then the following commutation and anti-commutation structures are satisfied
\begin{eqnarray}
 &&[{\cal H},{\cal H}] \in {\cal H}, \ [{\cal H}, {\cal A'}] \in {\cal A'},
 \ [{\cal A'},{\cal A'}] \in {\cal H}, \nonumber \\
 &&\{ {\cal H},{\cal H}\} \in {\cal A'},  \
 \{ {\cal H},{\cal A'}\} \in {\cal H}, \
  \{ {\cal A'},{\cal A'} \} \in {\cal A'}.
  \label{com-anticom}
\end{eqnarray}
We can see that $A'$ is another representation of $osp(1|32,R)$.
\par
  The definition of $gl(1|32,R)$ super Lie algebra is simply given by
  the following form of $33 \times 33$ supermatrix
  \begin{eqnarray}
    M = \left( \begin{array}{cc} m & \psi \\ i {\bar \phi} &
     v \end{array} \right),
   \end{eqnarray}
 where all components are real. 
The boson $m$ can be expanded similarly in terms of 11-dimensional 
$\gamma$-matrices as in (\ref{m-expansion-11}) but
all the coefficients  $u, \cdots ,u_{A_1 \cdots A_5}$ are real. 
Two fermions $\psi$
and $\phi$ are also real.
This matrix is decomposed into two parts as
\begin{equation}
 M = H + A,
\end{equation}
where
 \begin{eqnarray}
  H =
   \left( \begin{array}{cc} m_{h} & \psi_{1} \\ i {\bar \psi_{1}} & 0
   \end{array} \right),
   \nonumber \\
A=   \left( \begin{array}{cc} m_{a} & \psi_{2} \\  -i {\bar \psi_{2}}
   & v \end{array} \right).
  \end{eqnarray}
Here we have defined $\psi_{i}$ by
   \begin{eqnarray}
    \psi = \psi_{1} + \psi_{2}, \hspace{2mm} \phi = \psi_{1} - \psi_{2}
   \end{eqnarray}
and $m_h$ and $m_a$ are given in (\ref{mhma}) with
real coefficients.
%   and $m_{i}$ by
%   \begin{eqnarray}
%    & & m_{1} = u_{\mu_{1}} \Gamma^{\mu_{1}} + \frac{1}{2!} u_{\mu_{1}
%    \mu_{2}} \Gamma^{\mu_{1} \mu_{2}} + \frac{1}{5!} u_{\mu_{1} \cdots
%    \mu_{5}} \Gamma^{\mu_{1} \cdots \mu_{5}},  \nonumber \\
%    & & m_{2} = u {\bf 1} + \frac{1}{3!} u_{\mu_{1} \mu_{2} \mu_{3}}
%    \Gamma^{\mu_{1} \mu_{2} \mu_{3}} + \frac{1}{4!} u_{\mu_{1} \cdots
%    \mu_{4}} \Gamma^{\mu_{1} \cdots \mu_{4}}
%   \end{eqnarray}
%   with real coefficients.
$H$ is again an element of $osp(1|32,R)$ generators and $A$ is its
representation. 
 \par
 These two super Lie algebras
  $u(1|16,16)$ and $gl(1|32,R)$ are related as follows.
  A matrix $M=H + A'$ in $u(1|16,16)$ is mapped to a matrix
  in $gl(1|32,R)$ by $N =H+A$ where $A=iA'$ and  vice versa.
 Hence these two algebras are related by an analytic continuation.
\par
We now promote each real element of matrices to an $N \times N$ hermitian
matrix to make our model invariant under local 
$osp(1|32,R)$ symmetry.
  If we start from a set of  $osp(1|32,R)$ matrices ${\cal H}$
and make a tensor product with $u(N)$, the algebra does not close
within them because of the following relation:
    \begin{eqnarray}
      [ ({\cal H} \otimes {\bf H}) , ({\cal H} \otimes {\bf H}) ] &= &
     (\{ {\cal H}, {\cal H} \} \otimes [{\bf H},{\bf H}]) +
     ( [ {\cal H}, {\cal H}] \otimes \{ {\bf H}, {\bf H} \})
      \nonumber \\
        &=&
    ({\cal A}' \otimes {\bf A}) + ({\cal H} \otimes {\bf H}).
     \label{AZ52naivefail}     \end{eqnarray}
Here we have used (\ref{com-anticom}) and denoted ${\bf H}$ and
${\bf A}$ as hermitian and anti-hermitian matrices.
 In order for the algebra to close, it is necessary to include
  ${\cal A}' \otimes {\bf A}$. From the following relation
    \begin{eqnarray}
     [ ( {\cal A}' \otimes {\bf A}) , ({\cal A}' \otimes {\bf A}) ]& = &
     ( \{ {\cal A}', {\cal A}' \} \otimes [{\bf A},{\bf A}] ) +
     ( [ {\cal A}', {\cal A}'] \otimes \{ {\bf A}, {\bf A} \} )
     \nonumber \\
        &=&
     ( {\cal A}' \otimes {\bf A} ) + ( {\cal H} \otimes {\bf H} ) ,
     \label{AZ52close111}
   \end{eqnarray}
we can form a closed algebra by
combining 
 $({\cal H} \otimes {\bf H}) $ and
$ ({\cal A}' \otimes {\bf A})$  together.
For $N=1$ case, ${\bf H}$ and ${\bf A}$ are replaced by $1$ and $i$
respectively and it is nothing but $u(1|16,16)$ algebra 
we discussed before.
This is a reason why we need to enlarge $osp(1|32,R)$ to $u(1|16,16)$
for gauging the $osp(1|32,R)$ symmetry.
\par
Instead of promoting each element to a hermitian matrix, we can
make a closed algebra by restricting them to real matrices.
Since real matrices are closed under
commutators and anti-commutations, it is clear that
$({\cal H} + {\cal A}) \otimes gl(N,R) =gl(1|32,R)\otimes gl(N,R)$ forms
another closed algebra.
In this case, we have to embed the space-time into real matrices
instead of hermitian matrices.
%%%%%%%%%%%%%%%%%%%%%%%%%%%%%%%%%%%%%%%%%%%%%%%%%%%%%%%%55

 \subsection{Action and symmetries}
The action we consider is
 \begin{eqnarray}
   I &=& \frac{1}{g^{2}} Tr_{N \times N} \sum_{Q,R=1}^{33}
        ((\sum_{p=1}^{32}  {M_{p}}^{Q} {M_{Q}}^{R}
        {M_{R}}^{p})  -{M_{33}}^{Q} {M_{Q}}^{R}
        {M_{R}}^{33} ) = \frac{1}{g^{2}} Tr_{N \times N}( Str_{33
        \times 33} M^{3}) \nonumber \\
    &=& \frac{1}{g^{2}} \sum_{a,b,c=1}^{N^{2}} Str_{33 \times
        33}(M^{a} M^{b} M^{c} ) Tr_{N \times N} (t^{a} t^{b} t^{c}). 
\label{gaugedaction}
\end{eqnarray}
where $p = 1, \cdots, 32$ and $ Q, R = 1,  \cdots 33$. $M$ is a
supermatrix belonging to $u(1|16,16)$ or $gl(1|32,R)$ super Lie
algebra.
Each component ${M_{Q}}^{R}$ of the $33 \times 33$ supermatrix $M$ 
is promoted to an $N \times N$ matrix and can be expanded in terms
of Gell-Mann matrices:
\begin{equation}
 {M_{Q}}^{R} = \sum_{a=1}^{N^2} t^{a} {(M^{a})_{Q}}^{R}.
\end{equation}
 This action (\ref{gaugedaction}) is invariant under a 
tensor product of two gauge groups
    \begin{equation}
 M \Rightarrow M + [u, M],
     \end{equation}
where
\begin{equation}
 u \in gl(1|32,R) \otimes gl(N,R)  \textrm{ or } u(1|16,16) \otimes  u(N).
\end{equation}
Hence the action is invariant under local (or gauged) 
$u(1|16,16)$ or $gl(1|32,R)$
symmetry. That is,  the $u(1|16,16)$ symmetry and $u(N)$ symmetry
(or $gl(1|32,R)$ and $gl(N,R)$) are coupled.
Not only the bosonic but the fermionic
symmetries are also gauged. In this sense this action is considered as 
a matrix regularization of 11-dimensional supergravity 
if we can successfully treat this model.
In terms of the components of
$M = \left( \begin{array}{cc} m & \psi \\ i {\bar \phi} & v \end{array}
  \right),$ the action becomes
 \begin{equation}
 I = \frac{1}{g^{2}} Tr_{N \times N}  ( tr_{32 \times 32} (m^{3}) - 3i
 {\bar \phi} m \psi - 3i {\bar \phi} \psi v -
 v^{3}). \label{AZ53action} 
   \end{equation}
In both cases of $u(1|16,16)$ model and $gl(1|32,R)$ model, there are
$64=32+32$ (real) supercharges. 
The action is not invariant under the space-time translation
which was identified with a constant shift of bosonic fields 
in the case of IIB or $osp(1|32,R)$ model, and there are no
inhomogeneous supersymmetry in this gauged model. 
To extract space-time translation, we need
another interpretation different from the non-gauged model.
Here we adopt the Wigner-In{\"o}n{\"u} contraction of the $SO(10,1)$ symmetry
and identify $SO(9,1)$ rotation and space-time translation generators
with the $SO(10,1)$ generators.
In other words, we zoom in around the north pole of a 10-dimensional
sphere on which $SO(10,1)$ rotations are generated by $\Gamma^{AB}.$
\par
First let us consider a $gl(1|32,R)$-case.
To perform the Wigner-In{\"o}n{\"u} contraction systematically, 
it is convenient to add another term to the action
   \begin{eqnarray}
    I = \frac{1}{3} Tr_{N \times N} Str_{33 \times 33} ( M^{3}) -
    R^{2} Tr_{N \times N} Str_{33 \times 33} M. \label{AZ54veryaction}
   \end{eqnarray}
This action has a classical solution
  \begin{eqnarray}
  \langle M \rangle = \left( \begin{array}{cc} R \Gamma^{\sharp}
  \otimes {\bf 1}_{N \times N} & 0 \\ 0 & R \otimes {\bf 1}_{N \times N}
  \end{array} \right).
  \end{eqnarray}
We take a large $R$ limit, which is equivalent to zooming in around
the north pole.
\par
Similarly in the case of $u(1|16,16)$, we need to consider
a quintic
  action $I_{U(1|16,16)} = \frac{1}{5} Str(M^{5}) - R^{4} Str
  M$ in order to have the classical solution,
 \begin{equation}
\langle M \rangle =
  \left( \begin{array}{cc} R  \Gamma^{\sharp} \otimes {\bf 1}_{N
  \times N} & 0 \\ 0 & i R \otimes {\bf 1}_{N \times N} \end{array}
  \right).
\end{equation}
We focus on the $gl(1|32,R)$ type in this section, but
the following discussions of the Wigner-In{\"o}n{\"u} contraction are
essentially the same as in the $u(1|16,16)$ case.
\par
We expand the matrix $M$ around the above classical solution
$\langle M \rangle$:
   \begin{eqnarray}
    M =  \left
    ( \begin{array}{cc} m & \psi \\ i {\bar \phi} & v \end{array}
    \right) = \langle M \rangle + {\tilde M} = \left( \begin{array}{cc} R
    \Gamma^{\sharp} & 0 \\ 0 & R \end{array} \right) + \left
    ( \begin{array}{cc} {\tilde m} & \psi \\ i {\bar \phi} & {\tilde
    v} \end{array}  \right).
   \end{eqnarray}
The action becomes
   \begin{eqnarray}
    I = R( tr_{32 \times 32}({\tilde m}^{2} \Gamma^{\sharp}) - {\tilde 
    v}^{2} - i {\bar
    \phi} (1 + \Gamma^{\sharp}) \psi)  + \frac{1}{3} tr_{32 \times 32} 
    ({\tilde m}^{3}) - \frac{{\tilde v}^{3}}{3} -i ({\bar  \phi}
    {\tilde m} {\psi} + {\tilde v} {\bar \phi} \psi )
   \end{eqnarray}
up to a constant. In the next subsection,  we investigate the
model in the large $R$ limit. 

\subsection{Wigner-In{\"o}n{\"u} contraction and supersymmetry}
In the background proportional to $\Gamma^{\sharp}$, it
is convenient to decompose bosonic fields into even and odd
rank fields with respect to 10-dimensional $\gamma$-matrices:
\begin{equation}
 {\tilde m} = m_e + m_o,
\label{memo}
\end{equation}
where $m_e$ is given by
 \begin{eqnarray}
             m_{e} &=& Z {\bf 1} + W \Gamma^{\sharp} + \frac{1}{2}
             ( C_{\mu_{1} \mu_{2}} \Gamma^{\mu_{1} \mu_{2}} + 
            D_{\mu_{1} \mu_{2}}
             \Gamma^{\mu_{1} \mu_{2} \sharp} ) 
              \nonumber \\ 
            &+& \frac{1}{4!} ( G_{\mu_{1}
             \cdots \mu_{4}} \Gamma^{\mu_{1} \cdots \mu_{4}} +
             H_{\mu_{1} \cdots \mu_{4}} \Gamma^{\mu_{1} \cdots \mu_{4}
             \sharp}),  
            \end{eqnarray} 
and $m_o$ by
  \begin{eqnarray}
           m_{o} &=& \frac{1}{2} ( A^{(+)}_{\mu} \Gamma^{\mu} ( 1 +
           \Gamma^{\sharp}) + A^{(-)}_{\mu} \Gamma^{\mu} (1-
           \Gamma^{\sharp}) )              
          \nonumber \\ &+& 
           \frac{1}{2 \times 3!} (E^{(+)}_{\mu_{1} \mu_{2} \mu_{3}}
           \Gamma^{\mu_{1} \mu_{2} \mu_{3}} (1 + \Gamma^{\sharp} ) +
           E^{(-)}_{\mu_{1} \mu_{2} \mu_{3}} 
          \Gamma^{\mu_{1} \mu_{2} \mu_{3}} ( 1
           - \Gamma^{\sharp} ) ) \nonumber \\
           &+& \frac{1}{5!} (I^{(+)}_{\mu_{1} \cdots \mu_{5}}
           \Gamma^{\mu_{1} \cdots \mu_{5}} (1 + \Gamma^{\sharp}) +
           I^{(-)}_{\mu_{1} \cdots \mu_{5}} 
        \Gamma^{\mu_{1} \cdots \mu_{5}} (1
           - \Gamma^{\sharp}) ).
          \end{eqnarray}
We further decompose  the $m_o$ into 
 $m_{o}^{(\pm)}$ according to the 
$(\pm)$ in the above  decomposition: $m_o =m_{o}^{(+)} + m_{o}^{(-)}$.
 Fermionic fields are also decomposed according to their
       chiralities: $\psi_{L,R}= \frac{1 \pm \Gamma^{\sharp}}{2} \psi$.
  The action then becomes 
   \begin{eqnarray}
    I &=& R( tr_{32 \times 32}(m^{2}_{e} \Gamma^{\sharp}) - {\tilde
    v}^{2} - 2i {\bar \phi}_{R} \psi_{L} ) + tr_{32 \times 32}
    (\frac{1}{3} m^{3}_{e} + m_{e} m^{2}_{o} ) 
    \nonumber \\
    &-& i ( {\bar \phi}_{R} (m_{e} + {\tilde v}) \psi_{L}  + {\bar \phi}_{L}
    (m_{e} + {\tilde v}) \psi_{R} + {\bar \phi}_{L} m_{o} \psi_{L} + {\bar
    \phi}_{R} m_{o} \psi_{R} ) - \frac{1}{3} {\tilde v}^{3}.
 \label{AZ55kek}
   \end{eqnarray}
Since the quadratic term is proportional to $R$, we first
rescale ${\tilde v}, m_e, \phi_R$ and $\psi_L$ as $R^{-1/2}$.
Then, in order to make terms containing other fields 
such as $tr_{32 \times 32} (m_{e} m^{2}_{o})$ finite in the
large $R$ limit, we have to rescale the  other fields as $R^{1/4}$.
The rescalings are summarized as
  \begin{eqnarray}
  & &   m = R \Gamma^{\sharp} + {\tilde m} = R \Gamma^{\sharp} +
    R^{-\frac{1}{2}} m'_{e} + R^{\frac{1}{4}} m'_{o}, \hspace{3mm}
    v = R + {\tilde v} = R + R^{-\frac{1}{2}} v', \nonumber \\
  & & \psi = \psi_{L} + \psi_{R} = R^{-\frac{1}{2}} \psi'_{L} +
    R^{\frac{1}{4}} \psi'_{R} , \hspace{3mm} {\bar \phi} = {\bar
    \phi}_{L} + {\bar \phi}_{R} =
    R^{\frac{1}{4}} {\bar \phi}'_{L} + R^{-\frac{1}{2}} {\bar
    \phi}'_{R}.
   \end{eqnarray}
In terms of these rescaled fields,
we can rewrite the action, by dropping terms with a negative power of
$R$, as
   \begin{eqnarray}
    I &=& ( tr_{32 \times 32} (m'^{2}_{e} \Gamma^{\sharp}) - v'^{2} +
    tr_{32 \times 32} (m'_{e} m'^{2}_{o}) ) \nonumber \\
&& + i ( -2 {\bar \phi}'_{R} \psi'_{L} + {\bar \phi}'_{L}
    (m'_{e} + v') \psi'_{R} + {\bar \psi}'_{L} m'_{o} \psi'_{L} +
    {\bar \phi}'_{R} m'_{o} \psi'_{R} ).
   \end{eqnarray}
Since only the fields $v', m_e', \phi_R'$ and $\psi_L'$ have quadratic
terms, we may integrate them and obtain an effective action for the
other fields.
Before performing the integration,
let us first look at the supersymmetry structure in order to see
how we can obtain space-time supersymmetry in our model.
We can also see the above scalings are consistent with supersymmetries.
\par
The 10-dimensional space-time translation 
around the north pole is generated by
$\Gamma_{\mu \sharp}$. Since $R$ is interpreted as the radius
of $S^{10}$, space-time translation generator should be
identified with   
$P_{\mu} =  \frac{1}{R} \Gamma_{\mu  \sharp}$.
On the other hand, a commutator of two supercharges
       $Q_{\chi \epsilon} = \left( \begin{array}{cc} 0 & \chi \\ i
        {\bar \epsilon} & 0 \end{array} \right)$ and $Q_{\rho \eta} =
        \left( \begin{array}{cc} 0 & \rho \\ i {\bar \eta} & 0
        \end{array} \right) $ becomes
       \begin{eqnarray}
        [Q_{\chi \epsilon} , Q_{\rho \eta}] = \left( \begin{array}{cc}
        i (\chi {\bar \eta} - \rho {\bar \epsilon}) & 0 \\ 0 & i
        ({\bar \epsilon} \rho  - {\bar \eta} \chi) \end{array} \right),
       \end{eqnarray}
which contains the translation $P_{\mu}$ besides 
other $gl(32,R)$ bosonic generators. 
In this way, the homogeneous supersymmetry in gauged models 
is considered as 10-dimensional space-time supersymmetry.
\par
In addition to scaling the fields as above, we need to scale
gauge parameters of $gl(1|32,R)$. Writing the gauge parameter $h$  by
  \begin{eqnarray}
  h = \left( \begin{array}{cc} a & \chi \\ i {\bar \epsilon} & b
  \end{array} \right),
  \end{eqnarray}  the field $M$ is transformed as
   \begin{eqnarray}
   \delta {\tilde M} &=& [h,M] = [h, \langle M \rangle + {\tilde M}]
   \nonumber \\ 
   &=& \left( \begin{array}{cc} [a, {\tilde m} + R \Gamma^{\sharp}] + i
   ( \chi {\bar \phi} - \psi {\bar \epsilon}) & - ( {\tilde m} + R
   \Gamma^{\sharp}) \chi + a \psi - b \psi + \chi {\tilde v} \\ i {\bar
   \epsilon} ({\tilde m} + R \Gamma^{\sharp}) - i ( {\bar \phi} a +
   {\tilde v} {\bar
   \epsilon} ) + i b {\bar \phi} & i ({\bar \epsilon} \psi - {\bar
   \phi} \chi ) + [b, {\tilde v}] \end{array} \right), \label{AZ55subotr}
   \end{eqnarray}
where inhomogeneous terms come from $[h, \langle M \rangle]$.
The inhomogeneous term for $m_o$ should survive
after taking the large $R$ limit, since space-time translations for
 $A^{(\pm)}_{\mu}$ are included there.
Decomposing the bosonic gauge parameter $a$ into $a_o$ and $a_e$
similarly to (\ref{memo}),  
the inhomogeneous part of $\delta m_o$
is given by $\delta m_o = [a_o, R \Gamma^{\sharp}]$.
Since we have rescaled  $m_o = R^{1/4} m_o'$,
we should rescale $a_o$ as $R^{-3/4}$ so as
to make this inhomogeneous term finite in the large $R$ limit.
On the other hand, $SO(9,1)$ rotation generated by 
$\Gamma_{\mu \nu}$ is included in $a_e$ and it
transforms even (odd) rank fields into themselves.
The gauge parameter $a_e$ is, therefore, not necessary to be rescaled.
Similar arguments can be applied to supersymmetries and
we finally obtain the following rescalings
   \begin{eqnarray}
    h = \left( \begin{array}{cc} a'_{e} + R^{-\frac{3}{4}} a'_{o} &
    \chi'_{L} + R^{-\frac{3}{4}} \chi'_{R} \\ i ( R^{-\frac{3}{4}} {\bar
    \epsilon}'_{L} + {\bar \epsilon}'_{R}) & b' \end{array} \right).
    \label{AZ55chres}
   \end{eqnarray}
Under this gauge transformation, each field transforms 
in the large $R$ limit as
  \begin{eqnarray}
       \delta m'_{o} &=&
\underline{[ a_{o}, \Gamma^{\sharp}]} + [a'_{e}, m'_{o}]
     + i ( \chi'_{L} {\bar \phi}'_{L} - \psi'_{R} {\bar \epsilon}'_{R}
     ), \label{AZ55tr1964mo} \\
   \delta \psi'_{R} &=& \underline{2 \chi'_{R}} + ( a'_{e} \psi'_{R}  - b
       \psi'_{R} - (m'_{o} \chi'_{L}) ), \label{AZ55tr1964psir} \\
      \delta {\bar \phi}'_{L} &=& \underline{- 2 {\bar \epsilon}'_{L}}
           + ( ({\bar
         \epsilon}'_{R} m'_{o}) + b' {\bar \phi}'_{L} - {\bar
         \phi}'_{L} a'_{e}  ), \label{AZ55tr1964phil} \\
      \delta m'_{e} &=&  [a_{e}, m'_{e}] + [a_{o}, m'_{o}] + i
     ( \chi'_{L} {\bar \epsilon}'_{R} + \chi'_{R} {\bar \epsilon'_{L}}
     ) - i ( \psi'_{L} {\bar \epsilon}'_{R} + \chi'_{R} {\bar
     \epsilon}'_{L} ), \label{AZ55tr1964me} \\
      \delta \psi'_{L} &=& - ( m'_{o} \chi'_{R} + m'_{e} \chi_{L}) +
       (a_{o} \psi'_{R} + a_{e} \psi'_{L}) - b' \psi'_{L} + v'
       \chi'_{L}, \label{AZ55tr1964psil} \\
      \delta {\bar \phi}'_{R} &=& ( {\bar \epsilon}'_{R} m'_{e} +
      {\bar \epsilon}'_{L} m'_{o} ) + b' {\bar \phi}'_{R} - ({\bar
      \phi}'_{L} a'_{o} + {\bar \phi}'_{R} a'_{e} ) - v' {\bar
      \epsilon}'_{R}, \label{AZ55tr1964phir} \\
    \delta v' &=& i ( {\bar \epsilon}'_{R} \psi'_{L} + {\bar
         \epsilon}'_{L} m'_{o} ) - i ({\bar \phi}'_{R} \chi'_{L} +
         {\bar \phi}'_{L} \chi'_{R} ) + [b', v'].
 \label{AZ55tr1964v}
   \end{eqnarray}
The underlined terms are inhomogeneous transformations.
The other transformations are homogeneous and linear in fields.
As we have seen in the action, the fields that do not receive inhomogeneous
transformations, that is, $m'_e, v', \psi'_L$ and $\phi'_R$ 
contain quadratic terms and can be integrated out
by Gaussian integration.
An important point is that the transformations of the first three fields
$m_o', \psi_R'$ and $\phi_L'$ do not include the other fields in the
right hand side.
This means that these transformation rules 
 are not changed after integrating out the other fields, 
$m'_e, \psi'_L, \phi'_R$ and $v'$.
\par
Now let us obtain the effective action by integrating out
$v', m'_e, \psi'_L$ and $\phi'_R$.
The integration can be easily done and the effective action vanishes!
\begin{eqnarray}
W = - \frac{1}{4}  tr_{32 \times 32}( \Gamma^{\sharp} \{ m'^{2}_{o} + i
   (\psi'_{R} {\bar \phi}'_{L} ) \}^{2} ) - \frac{1}{4}  ({\bar
 \phi}'_{L} \psi_{R})^{2}  + \frac{i}{2} ( {\bar \phi}'_{L} m'^{2}_{o}
   \psi'_{R} )=0. \label{AZ55effectiveac}
   \end{eqnarray}
Here we have used
\begin{equation}
 tr_{32 \times 32} ( \Gamma^{\sharp}  {m'_{o}}^{4} ) =0.
\end{equation}
The reason for the vanishment of the effective action can be
understood from the symmetry point of view. 
Since the transformations for $m'_o, \psi'_R$ and $\phi'_L$
do not include the other integrated fields, they are not
changed after integration. Therefore the effective action 
must be invariant
under the same transformations, which include $U(N)$ dependent
shifts of them, not restricted to constant shifts.
The only  action invariant under such transformations is
a trivial one.
In this sense, this model is a topological model of fields
$m_o', \psi_R'$ and $\phi_L'$.
\par
Although the action vanishes, we go on to investigate the supersymmetry
structures.
If we decompose the boson fields $m'_o$ into $m_{o}^{'(\pm)}$,
we obtain the following transformations
   \begin{eqnarray}
    &&   \delta {m'_{o}}^{(+)} = - i (\psi'_{R} {\bar \epsilon}'_{R}),
\hspace{4mm}  \
 \delta {m'_{o}}^{(-)} =  i ( \chi'_{L} {\bar \phi}'_{L} ),
\nonumber \\
    &&  \delta \psi'_{R} = 2 \chi'_{R}  - ({m'_{o}}^{(+)} \chi'_{L}),
      \hspace{4mm}
      \delta {\bar \phi}'_{L} = - 2 {\bar \epsilon}'_{L} + ( {\bar
         \epsilon}'_{R} {m'_{o}}^{(-)} ), \label{AZ55susy1964}
   \end{eqnarray}
and we can see that left and right handed fields are decoupled.
We have two pairs of fields,
 ($m_o^{(+)}$ and $\psi_R$) and ($m_o^{(-)}$ and $\phi_L$).
This pairing is the same as that of $osp(1|32,R)$ model.
In this case, since the effective action vanishes we do not have the 
problem of compatibility with the pairing in the action. 
More explicitly in terms of ${A_{\mu}}^{(\pm)}$ fields,
the homogeneous supersymmetry transformations become
\begin{eqnarray}
  \delta^{(1)}_{\chi_{L}} A'^{(-)}_{i} = - \frac{i}{16} {\bar
    \phi}'_{L} \Gamma_{i} \chi_{L}, \hspace{10mm} 
     \delta^{(1)}_{\epsilon_{R}} A'^{(+)}_{i} =
   \frac{i}{16} {\bar \epsilon}_{R} \Gamma_{i} \psi'_{R},
\end{eqnarray}
which are the same as those of IIB matrix model.
On the other hand, transformations for fermions
are different from those of IIB matrix model. 
Instead of the commutators $[A^{(\pm)}_{\mu} ,A^{(\pm)}_{\nu} ]$, they
are proportional to a single $A^{(\pm)}_{\mu}$ and accordingly
supersymmetry parameters with opposite chirality.
Because of this reason,
it seems difficult to interpret IIB matrix model as a 
gauge fixed version of the gauged matrix model investigated here.
But such gauged models are interesting from various points of view,
especially the existence of local Lorentz invariance,
and it is worth further investigations. More analysis will be reported
elsewhere. 

\section{Conclusions and discussions}
In this paper, we have investigated several matrix models based on
super Lie algebra. First we have studied $osp(1|32,R)$ matrix model.
This model is considered as an 11-dimensional model and contains
twice as many fermionic degrees of freedom as IIB matrix model.
The model is invariant under global $osp(1|32,R)$ symmetry and 
$U(N)$ gauge symmetry. It is also invariant under a constant shift
of fields. Combining the $osp(1|32,R)$ and the constant shifts, we obtain
space-time algebras including space-time supersymmetries.
In this sense, this model is a natural generalization of 
IIB matrix model. Since this model has twice as many
fermions, we need to integrate half of the degrees of freedom.
We have given an identification of the fields in this model
with the fields in IIB matrix model from the view point of the
supersymmetry structures.
We have also discussed a possibility 
to induce IIB matrix model
by integrating out the unnecessary fields.
\par
In the latter half of this paper, we have studied the gauged matrix
models with local Lorentz symmetry.
 First we have shown that the unitary transformations
in noncommutative gauge theories contain much larger symmetries
than the ordinary gauge transformations. Especially, 
local coordinate transformations can be described within
this gauge transformations.
This is understandable from the D-brane point of view.
If a noncommutative gauge theory is considered as an effective
low-energy action for D-branes, the action should be invariant 
under coordinate transformations on the brane.
Under this transformation, all the fields $\psi$ and $A_{\mu}$ 
in gauge theory transform as scalars. 
More interestingly we have shown that if we expand those fields,
not only in terms of $\exp(i \hat{p}_{\mu} k_{\mu})$ but 
as a power series of  $\hat{p}_{\mu}$, we can obtain 
higher rank fields which transform as tensors under this coordinate
transformations. However, the original $SO(9,1)$ indices are
completely decoupled from the internal diffeomorphism.
\par
In the final section, we have considered a model with local
$SO(9,1)$ symmetry by extending the $osp(1|32,R)$ algebra to
$u(1|16,16)$ or $gl(1|32,R)$ super Lie algebras.
We have enhanced the global $osp(1|32,R)$ to local symmetries,
but lost the invariance under constant shifts of fields
and we need a different interpretation of space-time translation.
We have adopted the Wigner-In{\"o}n{\"u} contraction and extracted
10-dimensional space-time translation from $SO(10,1)$ rotations.
We  have then identified how to scale the fields in order to
obtain the correct 10-dimensional theory in the large
radius limit. Since this model contains
four times as many fermionic fields as IIB matrix model,
we need to integrate out half of them first and then restrict the
fermions further to be halved. 
But after integrating the first half of the fields,
the effective action was shown to vanish. 
This is because the resultant  action  should be invariant under
an arbitrary shift of the fields, not restricted to a constant shift.
We can interpret the final model as a topological model 
of IIB matrix model. This type of the topological model was studied in 
\cite{hiranokato}. 
It is also interesting to investigate such a possibility from the
gauged matrix model point of view.
%%%%%%%%%%%%%%%%%%%%%%%%%%%%%%%%%%%%%%%%%%%%%%%%%%%%%%%%%%%%%%%%%%
\appendix
\section{Supermatrices}
Here we briefly summarize our conventions of super Lie algebra. 
First the complex conjugate of Grassmann variables is defined by
\begin{equation}
 (\alpha \beta)^{\ast} = \beta^{\ast} \alpha^{\ast}.
\end{equation}
For a supermatrix $M =  \left( \begin{array}{cc} a & \beta \\ \gamma 
    & d \end{array} \right)  $ with bosonic matrices
$a, d$ and fermionic matrices $\beta, \gamma$, we define its transpose,
hermitian conjugate and complex conjugate as follows:
  \begin{eqnarray}
    {^{T}{M}} =  \left
    ( \begin{array}{cc} {^{T}{a}} & - {^{T}{\gamma}} \\ {^{T}{\beta}}
    & {^{T}{d}} \end{array} \right),
\hspace{5mm} 
M^{\dagger} = \left
    ( \begin{array}{cc} a^{\dagger} & \gamma^{\dagger} \\ \beta^{\dagger} & 
    d^{\dagger} \end{array} \right),
\hspace{5mm} 
M^{\ast} = ({^{T} M})^{\dagger}=
\left( \begin{array}{cc}
        a^{\ast} & \beta^{\ast} \\ - \gamma^{\ast} & d^{\ast}
        \end{array}  \right).
  \end{eqnarray}
From these definitions, we have the following relations
\begin{eqnarray}
  {^{T}{M}} = (M^{\ast})^{\dagger} \ne (M^{\dagger})^{\ast}, \ \
 M^{\dagger}= {^{T} (M^{\ast} )} \ne ({^{T}{M}})^{\ast}.
\end{eqnarray}
Note also that $ {^{T}{ ({^{T}{M}})}} \ne M$ while
$(M^{\dagger})^{\dagger}=M$ and$(M^{\ast})^{\ast}=M$.
For two supermatrices $M_1$ and $M_2$, we have the identities
${^{T}{(M_1 M_2)}} =  {^{T}{M_2}}  {^{T}{M_1}}$, 
$(M_1 M_2)^{\dagger} = M_2^{\dagger} M_1^{\dagger}$
and $(M_1 M_2)^{\ast} = M_1^{\ast} M_2^{\ast}$.
 
\section{Proof against a perturbative generation for  $W$ and 
$A_{\mu}^{(-)}$ propagators }
In this appendix we prove that it is impossible to generate
propagators for $W$ and $A_{\mu}^{(-)}$ fields 
through perturbative calculations in $osp(1|32,R)$
matrix model discussed in section 2. Of course, this proof 
does not exclude nonperturbative appearance of propagators for
them. 
 First we assign charges $(1,0,-1)$ to the bosonic fields
$(m_e, m_o^{(+)},m_o^{(-)})$ and $(0,1/2)$ to $(\psi_L, \psi_R).$
As we can see from the action (\ref{osp-action-eo}), every three point 
vertex has charge $3,3/2$ or $0$. Similarly in the background of
(\ref{classical-background}), propagators appear at tree level for 
$\langle m_e m_o^{(-)} \rangle$ and $\langle \psi_L \bar{\psi}_L\rangle$
and they have charges $1$ and $0$ respectively.
On the other hand, two point function $\langle WW \rangle$
which is included in $\langle m_e m_e \rangle$  or
$\langle A_{\mu}^{(-)} A_{\nu}^{(-)} \rangle$ in 
$\langle m_o^{(-)} m_o^{(-)} \rangle$ has charge $2$ or $-2$ 
respectively. Hence it is clearly impossible to
generate these two point functions perturbatively no matter
how we combine the above vertices and tree level propagators.
%%%%%%%%%%%%%%%%%%%%%%%%%%%%%%%%%%%%%%%%%%%%%%%%%%%%%%%%%%%%%%%%%%%%%%%%%%%%%
%\begin{center} \begin{large}
%Acknowledgments
%\end{large} \end{center}

\end{document}